\documentclass[english,final,leqno,onefignum,onetabnum]{siamltexmm}
\usepackage[T1]{fontenc}
\usepackage[latin9]{inputenc}
\usepackage{array}
\usepackage{booktabs}
\usepackage{textcomp}
\usepackage{multirow}
\usepackage{amsmath}
\usepackage{amssymb}
\usepackage{graphicx}
\usepackage{setspace}
\usepackage{esint}
\usepackage{babel}

\title{A Flexible Uncertainty Quantification Framework for General Multi-Physics
Systems} 

\author{A.~Mittal\footnotemark[2]
\and X.~Chen\footnotemark[1] \footnotemark[4]
\and C.~H. Tong\footnotemark[1]
\and G. Iaccarino\footnotemark[3]}

\begin{document}
\maketitle
\newcommand{\slugmaster}{%
\slugger{juq}{xxxx}{xx}{x}{x--x}}

\renewcommand{\thefootnote}{\fnsymbol{footnote}}

\footnotetext[1]{Center for Applied Scientific Computing, Lawrence Livermore National Laboratory, Livermore, CA 94550.}
\footnotetext[2]{Institute for Computational and Mathematical Engineering, Stanford University, Stanford, CA 94305.}
\footnotetext[3]{Mechanical Engineering, Stanford University, Stanford, CA 94305.}
\footnotetext[4]{Corresponding author (Email: \email{chen73@llnl.gov})}
\renewcommand{\thefootnote}{\arabic{footnote}}

\begin{abstract}
We present a ``module-based hybrid'' Uncertainty Quantification (UQ)
framework for general nonlinear multi-physics simulation. The proposed
methodology, introduced in [\hyperlink{ref1}{1}], supports the independent development of each \emph{stochastic}
linear or nonlinear physics module equipped with the most suitable probabilistic
UQ method: non-intrusive, semi-intrusive or intrusive; and provides
a generic framework to couple these stochastic simulation components.
Moreover, the methodology is illustrated using a
common ``global'' uncertainty representation scheme based on generalized
polynomial chaos (gPC) expansions of inputs and outputs. By using
thermally-driven cavity flow as the multi-physics model problem, we demonstrate
the utility of our framework and report the computational gains achieved.
\end{abstract}



\pagestyle{myheadings}
\thispagestyle{plain}
\markboth{A.~MITTAL, X.~CHEN, C. H. ~TONG AND G. IACCARINO}{STOCHASTIC MODELING OF MULTI-PHYSICS SYSTEMS}

\section{Introduction}

The discipline of Uncertainty Quantification (UQ) seeks to develop and apply rigorous methodologies
to determine uncertainties associated the modeling and simulation
of physical processes. The goal is to estimate the probabilistic
variations and associated confidence intervals in the quantity of
interest resulting from all relevant sources of uncertainty (uncertainty analysis)
and to rank the contribution of individual sources of uncertainties
(sensitivity analysis). Advances in mathematical/statistical techniques
and the availability of high performance computers in recent years
have provided an unprecedented opportunity to undertake the computationally
intensive task of ``model predictions with confidence'' in complex
multi-physics applications. 

Broadly speaking, UQ approaches can be categorized as either non-intrusive
or intrusive. Non-intrusive methods such as Monte-Carlo (MC) generate
a statistical description of the model output by first drawing random samples
from a given probability distribution, running deterministic simulations
with those samples, and finally computing the output statistics
and/or sensitivities. The main advantages of these methods is the
simplicity of implementation using deterministic simulation codes
and the embarrassingly parallel computing possibilities. However,
these methods suffer from  slow convergence rate. Many
alternative random sampling designs such as quasi-Monte Carlo [\hyperlink{ref2}{2}],
Latin Hypercube [\hyperlink{ref3}{3}] and importance sampling [\hyperlink{ref4}{4}] have been
proposed. 

Intrusive methods, on the other hand, generally require a re-formulation
of deterministic models. A popular class of intrusive methods is the
stochastic Galerkin method based on generalized Polynomial Chaos (gPC)
expansion [\hyperlink{ref5}{5}, \hyperlink{ref6}{6}, \hyperlink{ref7}{7}]. gPC has been used successfully in many applications
such as solid mechanics [\hyperlink{ref5}{5}], transport in heterogeneous media
[\hyperlink{ref8}{8}], fluid mechanics [\hyperlink{ref9}{9}, \hyperlink{ref10}{10}], combustion [\hyperlink{ref11}{11}], etc. The
advantage of intrusive gPC-based methods is that they may have excellent
convergence properties when compared to MC-based methods [\hyperlink{ref12}{12}, \hyperlink{ref13}{13}].
However, the rapidly increasing complexity and fidelity of multi-physics
models have limited the popularity of intrusive methods. A major reason
is that implementation of such methods requires extensive modifications
to existing deterministic codes, a task that may be too cumbersome
and time-consuming, especially for complex and nonlinearly coupled
multi-physics models. The size of the coupled system arising from spatio-temporal
discretizations may become so large that the implementation of any
further stochastic projection schemes, such as Galerkin (SGS), become
computationally intractable. Moreover, additional challenges in implementing
intrusive methods yet remain unresolved for complex unsteady applications,
such as turbulent flow and highly nonlinear transient problems. For
a detailed review of intrusive gPC-based uncertainty propagation for
CFD applications, we refer to [\hyperlink{ref9}{9}, \hyperlink{ref14}{14}].

To overcome some of these limitations, non-intrusive gPC methods have
been proposed as viable alternatives (c.f. [\hyperlink{ref15}{15}]). These methods
use either regression or quadrature techniques to estimate the coefficients
of the gPC expansions and can typically exhibit improved convergence behavior
[\hyperlink{ref16}{16}] over MC-based methods. In regression-based techniques, oversampling is often required
to compute accurate solutions, while in quadrature-based gPC methods,
MC-based random sampling is replaced with evaluations corresponding
to numerical integration rules (often with very strict constraints). These limitations can render these
methods unattractive in practice. Moreover, both intrusive and non-intrusive
gPC methods suffer from the so-called curse-of-dimensionality, where
the computational effort required grows exponentially with the number
of independent sources of uncertainty. Recent developments (stochastic
collocation [\hyperlink{ref17}{17}, \hyperlink{ref18}{18}], low-rank approximations [\hyperlink{ref19}{19}], radial
basis functions [\hyperlink{ref20}{20}], Pade-Legendre approaches [\hyperlink{ref21}{21}], and response
surface reconstruction [\hyperlink{ref22}{22}]) have demonstrated how the mathematical
structure of a model and the regularity of the solutions can be exploited
to achieve superlinear convergence [\hyperlink{ref18}{18}]. Various ``hybrid''
approaches that combine intrusive and non-intrusive methods have also
been recently proposed. Examples include the multi-state procedure
[\hyperlink{ref23}{23}], the mixed aleatory/epistemic representation approach [\hyperlink{ref24}{24}]
and the domain hybridization method [\hyperlink{ref25}{25}], the development of which
was driven by the need to couple two different descriptions of turbulent
flows.

In this article, we propose an alternative hybrid (or partially intrusive)
framework for uncertainty propagation in modular multi-physics simulations.
Such a framework was initially proposed for linear multi-physics problems
in a previous article [\hyperlink{ref1}{1}], and we aim to tackle general nonlinear
applications in this work. As motivated in [\hyperlink{ref1}{1}], the proposed hybrid
framework can blend UQ methods, intrusive or non-intrusive that are
best suited or available for each individual solver module, and seamlessly
``glue'' them together to facilitate global uncertainty/sensitivity
propagation. To formalize the notion of a modular solution framework,
we consider an algebraic system of equations that represents an $m-$component
multi-physics system as follows. \hypertarget{eq11}{}
\begin{equation}
\boldsymbol{f}_{i}\left(\boldsymbol{u}_{i},\boldsymbol{u}_{1},\ldots,\boldsymbol{u}_{i-1},\boldsymbol{u}_{i+1},\ldots,\boldsymbol{u}_{m},\boldsymbol{\xi}_{i}\right)=\boldsymbol{0},\ 1\leq i\leq m,
\end{equation}
where, $\boldsymbol{u}_{i}\in\mathbb{R}^{n_{i}}$ and $\boldsymbol{\xi}_{i}\in\mathbb{R}^{s_{i}}$
correspond to the solution field and input parameters in the $i$-th
component respectively. A differential system of equations can be
reduced to the algebraic form in Eq. \hyperlink{eq11}{1.1} by appropriate discretization schemes
in space and time. By implementing an iterative (staggered) solution
approach [\hyperlink{ref26}{26}], existing (legacy) solvers for each module $i$
can be leveraged as independent computational kernels to solve Eq. \hyperlink{eq11}{1.1}.
At iteration $\ell\geq0$ and module $1\leq i\leq m$, we have
\hypertarget{eq12}{}
\begin{equation}
\boldsymbol{u}_{i}^{\ell+1}\left(\boldsymbol{\xi}_{1},\ldots,\boldsymbol{\xi}_{m}\right)=\boldsymbol{m}_{i}\left(\boldsymbol{u}_{1}^{\ell}\left(\boldsymbol{\xi}_{1},\ldots,\boldsymbol{\xi}_{m}\right),\ldots,\boldsymbol{u}_{m}^{\ell}\left(\boldsymbol{\xi}_{1},\ldots,\boldsymbol{\xi}_{m}\right),\boldsymbol{\xi}_{i}\right).
\end{equation}

The iterations shown in Eq. \hyperlink{eq12}{1.2} are performed until each the
norm of each solution update $\left\Vert \boldsymbol{u}_{i}^{\ell+1}-\boldsymbol{u}_{i}^{\ell}\right\Vert $
falls below a prescribed tolerance. The solution from the previous
iteration step $\boldsymbol{u}_{i}^{\ell}$ may enter into the $i$-th
module as an initial guess and therefore, has been included as an
argument in the module operator $\boldsymbol{m}_{i}$. Therefore,
compared to a monolithic approach (fully-coupled solvers) solving
Eq. \hyperlink{eq11}{1.1}, the partitioned solution approach only requires the
construction of an additional iteration controller which allows individual
single-physics modules to be updated and replaced independently. From
practical considerations, this approach enables an attractive
``plug-and-play'' framework for developing multi-physics simulation
software. Due to modeling and measurement errors, exact values of
the input parameters in Eq. \hyperlink{eq11}{1.1} are usually not precisely known
and therefore, we model these quantities as random variables (with
a prescribed statistical description). The goal is to compute uncertainties
in the quantities of interest in the form of probability distributions,
statistics, and sensitivity information. All these tasks can be efficiently
achieved within the proposed hybrid framework using gPC methods. 

The remainder of this article is devoted to the description of the proposed
module-based hybrid UQ framework. In $\S$\hyperlink{sec2}{2}, we provide
a brief overview of gPC based intrusive methods, non-intrusive methods,
and semi-intrusive methods that exploit additional derivative information.
In $\S$\hyperlink{sec3}{3}, we detail the module-based hybrid computational framework
associated with modular gPC representations. In $\S$\hyperlink{sec4}{4}, we demonstrate
an implementation our proposed framework thermally driven cavity flows
as the numerical multi-physics example.

\hypertarget{sec2}{}
\section{Overview of gPC based UQ methods}

Propagating uncertainty and sensitivity information using gPC is a
popular choice in cases where the solution is expected to behave regularly
in the input stochastic space. We begin this review section by introducing
some definitions that will be used throughout the article. Let random
inputs $\boldsymbol{\xi}_{1},\ldots,\boldsymbol{\xi}_{m}$ belong to a
complete probability space $\left(\Xi,\mathcal{B}\left(\Xi\right),\mathcal{P}\right)$,
where $\Xi$ is the sample space (set of outcomes), $\mathcal{B}$
denotes the Borel measure and $\mathcal{P}:\Xi\rightarrow\left[0,1\right]$
is a probability measure. We assume that the constituent random scalar
components $\xi_{ij}:1\leq i\leq m,1\leq j\leq s_{i}$ are independent
and belong to a probability space $\left(\Xi_{ij},\mathcal{B}\left(\Xi_{ij}\right),\mathcal{P}_{ij}\right)$,
where $\Xi_{ij}\subseteq\mathbb{R}$. Moreover, we define $\Xi_{i}=\bigcup_{j=1}^{s_{i}}\Xi_{ij}$.
Furthermore, let $s=\sum_{i=1}^{m}s_{i}$ denote the dimension
of $\Xi$. If all the moments of $\mathcal{P}_{ij}$ are finite,
then a corresponding set of orthonormal polynomials [\hyperlink{ref27}{27}] can be
defined as follows. 
\begin{equation}
\left\{ \psi_{ij}^{k}:k\geq0\right\} :\int_{\mathbb{R}}\psi_{ij}^{k}\left(\xi\right)\psi_{ij}^{l}\left(\xi\right)d\mathcal{P}_{ij}\left(\xi\right)=\delta_{kl}.
\end{equation}

The orthonormality condition gives rise to a three term recurrence
property of the polynomials as follows. $\forall \xi\in\Xi_{ij},k\geq0$,
\begin{equation}
\xi\psi_{ij}^{k}\left(\xi\right)=\sqrt{\beta_{k+1}}\psi_{ij}^{k+1}\left(\xi\right)+\alpha_{k}\psi_{ij}^{k}\left(\xi\right)+\sqrt{\beta_{k}}\psi_{ij}^{k-1}\left(\xi\right)
\end{equation}
with $\psi_{ij}^{-1}=0$. The Chebyshev algorithm [\hyperlink{ref27}{27}] can be
used to obtain the coefficients of the recurrence relation (4) from
the raw moments of $\mathcal{P}_{ij}$. If $\mathcal{P}_{ij}$ is
a well known probability measure, the coefficients can be analytically
obtained from the Weiner-Askey tables [\hyperlink{ref28}{28}]. Once the univariate
polynomials are constructed, their multivariate extensions can be
naturally constructed by tensorization. We define the component basis
polynomials as follows.

\begin{equation}
\left\{ \psi_{i}^{\boldsymbol{j}}:\boldsymbol{j}=\left(j_{1}\ldots j_{s_{i}}\right)\in\mathbb{N}_{0}^{s_{i}}\right\} :\psi_{i}^{\boldsymbol{j}}\left(\boldsymbol{\xi}_{i}\right)=\psi_{i}^{\boldsymbol{j}}\left(\xi_{i1},\ldots,\xi_{is_{i}}\right)=\prod_{k=1}^{s_{i}}\psi_{ik}^{j_{k}}\left(\xi_{ik}\right).
\end{equation}

Using the component basis polynomials, we define the global basis
polynomials as follows.
\begin{equation}
\left\{ \psi^{\boldsymbol{j}}:\boldsymbol{j}=\left(\boldsymbol{j}_{1}\ldots\boldsymbol{j}_{m}\right)\in\mathbb{N}_{0}^{s_{1}}\times\cdots\times\mathbb{N}_{0}^{s_{m}}\right\} :\psi^{\boldsymbol{j}}\left(\boldsymbol{\xi}_{1},\ldots,\boldsymbol{\xi}_{m}\right)=\prod_{k=1}^{m}\psi_{k}^{\boldsymbol{j}_{k}}\left(\boldsymbol{\xi}_{k}\right).
\end{equation}

Assuming that the solution fields in Eq. \hyperlink{eq11}{1.1} are second order
random variables, we can define them in terms of an infinite series
of the respective orthonormal polynomials as follows. $\forall1\leq i\leq m,$ \hypertarget{eq25}{}
\begin{align}
\boldsymbol{u}_{i}\left(\boldsymbol{\xi}_{1},\ldots,\boldsymbol{\xi}_{m}\right) & =\sum_{\left|\boldsymbol{j}\right|\geq0}\hat{u}_{i}^{\boldsymbol{j}}\psi^{\boldsymbol{j}}\left(\boldsymbol{\xi}_{1},\ldots,\boldsymbol{\xi}_{m}\right).
\end{align}

Defining a total order $p\geq0$, we can truncate the infinite
series in Eq. \hyperlink{eq25}{2.5} as follows.

\begin{equation}
\boldsymbol{u}_{i}^{p}\left(\boldsymbol{\xi}_{1},\ldots,\boldsymbol{\xi}_{m}\right)\approx\sum_{\left|\boldsymbol{j}\right|=0}^{p}\hat{u}_{i}^{\boldsymbol{j}}\psi^{\boldsymbol{j}}\left(\boldsymbol{\xi}_{1},\ldots,\boldsymbol{\xi}_{m}\right).
\end{equation}

As stated by the Cameron-Martin theorem [\hyperlink{ref29}{29}], if the solution
fields are sufficiently regular functions of the random variables,
then the truncated approximation $\boldsymbol{u}_{i}^{p}$ converges
exponentially to $\boldsymbol{u}_{i}$, in the $\mathcal{L}_{2}-$sense,
as $p\rightarrow\infty$. The coefficients of the expansion in equation
(6) are known as the global gPC coefficients. The gPC approximations
can also be defined using a single-index and matrix-vector product
form as follows. 
\begin{align}
\boldsymbol{u}_{i}^{p}\left(\boldsymbol{\xi}_{1},\ldots,\boldsymbol{\xi}_{m}\right) & =\sum_{j=0}^{P}\hat{\boldsymbol{u}}^{j}\psi^{j}\left(\boldsymbol{\xi}_{1},\ldots,\boldsymbol{\xi}_{m}\right)=\hat{\boldsymbol{U}}_{i}\boldsymbol{\psi}\left(\boldsymbol{\xi}_{1},\ldots,\boldsymbol{\xi}_{m}\right),
\end{align}
where $\hat{\boldsymbol{U}}_{i}=\hat{\boldsymbol{U}}_{i}^{p}=\left[\begin{array}{ccc}
\hat{\boldsymbol{u}}_{i}^{0} & \cdots & \hat{\boldsymbol{u}}_{i}^{P}\end{array}\right]$ denotes the gPC-based coefficient matrix,  $\boldsymbol{\psi}=\boldsymbol{\psi}^{p}=\left[\begin{array}{ccc}
\psi^{0} & \cdots & \psi^{P}\end{array}\right]^\mathbf{T}$ denotes the basis vector and $P+1={p+s \choose p}$ is the cardinality
of the basis. 

The gPC coefficients have a simple relationship to the first two moments
of the solutions, which can be written as follows. 
\begin{align}
\mathbb{E}\left(\boldsymbol{u}_{i}\right) & \approx\mathbb{E}\left(\boldsymbol{u}_{i}^{p}\right)=\hat{u}_{i}^{0},\nonumber \\
\mathrm{Cov}\left(\boldsymbol{u}_{i},\boldsymbol{u}_{i}\right) & \approx\mathrm{Cov}\left(\boldsymbol{u}_{i}^{p},\boldsymbol{u}_{i}^{p}\right)=\sum_{j=1}^{P}\hat{u}_{i}^{j}\left(\hat{u}_{i}^{j}\right)^\mathbf{T}.
\end{align}

Moreover, since polynomials are orders of magnitude cheaper to compute
in comparison to solving the multi-physics system in Eq. \hyperlink{eq11}{1.1},
higher order statistics of the solution fields can be subsequently
estimated with exhaustive MC sampling. Similarly, probability distributions
of related quantities of interest can be accurately estimated using
the kernel density (KDE) method [\hyperlink{ref30}{30}]. Furthermore, global sensitivity
indices using the ANOVA [\hyperlink{ref31}{31}] method can also be directly obtained
from the gPC coefficients.

We will now describe how to propagate the gPC coefficients using non-intrusive,
semi-intrusive methods and intrusive gPC-based methods. For notational
simplicity, the methods will be discussed in the context of a single-physics
model, which represents a single component of a multi-physics model,
and can be formulated as follows.\hypertarget{eq29}{} 
\begin{equation}
\boldsymbol{f}\left(\boldsymbol{u},\boldsymbol{v},\boldsymbol{\xi}\right)=\boldsymbol{0}:\boldsymbol{f},\boldsymbol{u}\in\mathbb{R}^{n},\boldsymbol{v}\in\mathbb{R}^{\tilde{n}},\boldsymbol{\xi}\in\Xi\subseteq\mathbb{R}^{s},
\end{equation}
where $\boldsymbol{u}$ is the solution variable, $\boldsymbol{v}$
is the auxiliary or coupling variable and $\boldsymbol{\xi}$ is the
(random) input parameter. The objective here is to compute the solution
gPC coefficient matrix $\hat{\boldsymbol{U}}$ given $\hat{\boldsymbol{V}}:\boldsymbol{v}\approx\hat{\boldsymbol{V}}\boldsymbol{\psi}$.

\subsection{Non-intrusive methods}

Non-intrusive gPC methods are based on reusing a deterministic solver
which can be executed for various input parameter values. For a fixed
$Q-$sized sampling design $\left\{ \boldsymbol{\xi}^{\left(j\right)}\in\Xi\right\} _{j=1}^{Q}$,
we precompute the basis vector samples $\left\{ \boldsymbol{\psi}^{\left(j\right)}=\boldsymbol{\psi}\left(\boldsymbol{\xi}^{\left(j\right)}\right)\right\} _{j=1}^{Q}$
and construct $\boldsymbol{\Psi}=\boldsymbol{\Psi}^{p,Q}=\left[\begin{array}{ccc}
\boldsymbol{\psi}^{\left(1\right)} & \cdots & \boldsymbol{\psi}^{\left(Q\right)}\end{array}\right]$, known as the Fisher matrix [\hyperlink{ref32}{32}]. Subsequently, we construct
the solution sample matrix $\boldsymbol{U}=\boldsymbol{U}^{Q}=\left[\begin{array}{ccc}
\boldsymbol{u}^{\left(1\right)} & \cdots & \boldsymbol{u}^{\left(Q\right)}\end{array}\right]:\forall1\leq j\leq Q,\boldsymbol{f}\left(\boldsymbol{u}^{\left(j\right)},\hat{\boldsymbol{V}}\boldsymbol{\psi}^{\left(j\right)},\boldsymbol{\xi}^{\left(j\right)}\right)=\boldsymbol{0}$. Then, either of the following methods can be used to compute \textbf{$\hat{\boldsymbol{U}}$}.

\subsubsection{Polynomial regression}

In this method, $\hat{\boldsymbol{U}}$ is the analytical solution
of a least-squares minimization problem, as follows. \hypertarget{eq210}{} 
\begin{equation}
\hat{\boldsymbol{U}}=\arg\min_{\hat{\boldsymbol{Y}}\in\mathbb{R}^{n\times\left(P+1\right)}}\left\Vert \boldsymbol{U}-\hat{\boldsymbol{Y}}\boldsymbol{\Psi}\right\Vert _{F}=\boldsymbol{U}\boldsymbol{\Psi}^\mathbf{T}\left(\boldsymbol{\Psi}\boldsymbol{\Psi}^\mathbf{T}\right)^{-1},
\end{equation}
where $\left\Vert \cdot\right\Vert _{F}$ denotes the Frobenius norm.
A proof of Eq. \hyperlink{eq210}{2.10} has been provided in Lemma \hyperlink{lemA1}{A1}, in Appendix
\hyperlink{appA}{A}

To ensure that $\boldsymbol{\Psi}$ is nonsingular, we have the lower
bound $Q_{\min}=P+1$. Moreover, to ensure stability and that the
condition number of $\boldsymbol{\Psi}$ remains reasonably low, a
sample size of twice the lower bound is typically enforced.

\subsubsection{Pseudospectral approximation}

Alternatively, we can compute $\hat{\boldsymbol{U}}$ using numerical
integration (quadrature) methods [\hyperlink{ref33}{33}]. If $\left\{ \left(\boldsymbol{\xi}^{\left(j\right)},w^{\left(j\right)}\right)\right\} _{j=1}^{Q}$
denotes a quadrature rule in $\Xi$ , we can approximate the gPC
coefficient matrix as follows.\hypertarget{eq211}{}
\begin{equation}
\hat{\boldsymbol{U}}=\int_{\Xi}\boldsymbol{u}\left(\boldsymbol{\xi}\right)\left(\boldsymbol{\psi}\left(\boldsymbol{\xi}\right)\right)^\mathbf{T}d\mathcal{P}\left(\boldsymbol{\xi}\right)\approx\sum_{k=1}^{Q}\boldsymbol{u}^{\left(j\right)}\boldsymbol{\psi}^{\left(j\right)}\left(\boldsymbol{\xi}^{\left(k\right)}\right)w^{\left(j\right)}=\boldsymbol{U}\boldsymbol{W}\boldsymbol{\Psi}^\mathbf{T},
\end{equation}
where $\boldsymbol{W}=\boldsymbol{W}^{Q}=diag\left\{ w^{\left(1\right)},\ldots,w^{\left(Q\right)}\right\} $.
If the level of the quadrature rule is $\geq p$, the pseudospectral
approximation in Eq. \hyperlink{eq211}{2.11}  is equivalent to a weighted regression
method using the sample matrix $\boldsymbol{U}$. Lemma \hyperlink{lemA2}{A2}  in Appendix
\hyperlink{appA}{A} proves this equivalence.

\subsection{Semi-intrusive methods}

Semi-intrusive gPC methods are based on extracting additional stochastic
information from the model with minimal modifications to the deterministic
solver. A popular choice is to extract the first derivatives (gradients)
of the solution with respect to the input parameters. In the context
of Eq. \hyperlink{eq29}{2.9} , the gradients can be obtained by using the property
of the total derivative of $\boldsymbol{f}$ with respect to the input
parameters $\boldsymbol{\xi}$ as follows.
\begin{align}
 & \frac{\partial\boldsymbol{f}}{\partial\boldsymbol{\xi}}+\left(\frac{\partial\boldsymbol{f}}{\partial\boldsymbol{u}}\right)\frac{\partial\boldsymbol{u}}{\partial\boldsymbol{\xi}}+\left(\frac{\partial\boldsymbol{f}}{\partial\boldsymbol{v}}\right)\frac{\partial\boldsymbol{v}}{\partial\boldsymbol{\xi}}=\boldsymbol{0}\\
\Rightarrow & \frac{\partial\boldsymbol{u}}{\partial\boldsymbol{\xi}}=-\left(\frac{\partial\boldsymbol{f}}{\partial\boldsymbol{u}}\right)^{-1}\left(\frac{\partial\boldsymbol{f}}{\partial\boldsymbol{\xi}}+\left(\frac{\partial\boldsymbol{f}}{\partial\boldsymbol{v}}\right)\frac{\partial\boldsymbol{v}}{\partial\boldsymbol{\xi}}\right).
\end{align}

Therefore, the solver would need to be modified slightly to obtain
$\frac{\partial\boldsymbol{f}}{\partial\boldsymbol{u}},$ $\frac{\partial\boldsymbol{f}}{\partial\boldsymbol{v}}$
and $\frac{\partial\boldsymbol{f}}{\partial\boldsymbol{\xi}}$. If,
for instance, Newton's method is used to solve Eq. \hyperlink{eq29}{2.9}, we can
simply reuse the Jacobian $\frac{\partial\boldsymbol{f}}{\partial\boldsymbol{u}}$
at the last iteration, Moreover, the derivative of $\boldsymbol{v}$
is approximated as

\begin{align}
\frac{\partial\boldsymbol{v}}{\partial\boldsymbol{\xi}}\left(\boldsymbol{\xi}\right) & \approx\hat{\boldsymbol{V}}\frac{\partial\boldsymbol{\psi}}{\partial\boldsymbol{\xi}}\left(\boldsymbol{\xi}\right).
\end{align}

Following a similar approach for regression without derivatives, for
a fixed $Q-$sized sampling design $\left\{ \boldsymbol{\xi}^{\left(j\right)}\in\Xi\right\} _{j=1}^{Q}$,
we precompute samples of the basis vectors and their derivatives $\left\{ \tilde{\boldsymbol{\psi}}^{\left(j\right)}=\left[\begin{array}{cc}
\boldsymbol{\psi}^{\left(j\right)} & \frac{\partial\boldsymbol{\psi}^{\left(j\right)}}{\partial\boldsymbol{\xi}}\end{array}\right]\right\} _{j=1}^{Q}$, and construct the modified Fisher matrix that can be written as $\tilde{\boldsymbol{\Psi}}=\tilde{\boldsymbol{\Psi}}^{p,Q}\left[\begin{array}{ccc}
\tilde{\boldsymbol{\psi}}^{\left(1\right)} & \cdots & \tilde{\boldsymbol{\psi}}^{\left(Q\right)}\end{array}\right]$. Subsequently, we run the modified solver $Q$ times and construct
the solution and derivative sample matrix $\tilde{\boldsymbol{U}}=\left[\begin{array}{ccccc}
\boldsymbol{u}^{\left(1\right)} & \frac{\partial\boldsymbol{u}^{\left(1\right)}}{\partial\boldsymbol{\xi}} & \cdots & \boldsymbol{u}^{\left(Q\right)} & \frac{\partial\boldsymbol{u}^{\left(Q\right)}}{\partial\boldsymbol{\xi}}\end{array}\right]$.

To compute the gPC coefficient matrix $\hat{\boldsymbol{U}}$, we
can use the following analytical solution (Lemma \hyperlink{lemA1}{A1}) of the least-squares
minimization problem.
\begin{equation}
\hat{\boldsymbol{U}}=\tilde{\boldsymbol{U}}\tilde{\boldsymbol{\Psi}}^\mathbf{T}\left(\tilde{\boldsymbol{\Psi}}\tilde{\boldsymbol{\Psi}}^\mathbf{T}\right)^{-1},
\end{equation}
To ensure that $\tilde{\boldsymbol{\Psi}}$ is nonsingular, the lower
bound on the sample size: $Q_{\min}=\left\lceil \frac{P+1}{s+1}\right\rceil $.
For numerical stability, a sample size of twice the lower bound is
usually enforced. With the additional first derivative information,
the sample size would therefore, be $s$ times smaller than in non-intrusive
case without derivative information. Moreover, the additional cost
of obtaining the derivatives is an additional $s$ Newton solves,
implying that the ratio of computational costs between the semi-intrusive
and non-intrusive regression methods would be $\frac{1}{s+1}\left(1+\frac{s}{k}\right)$,
where $k$ is the number of iterations needed to converge to the solution.
When $k\gg s$, this ratio is $\approx\frac{1}{s+1}$.

\subsection{Intrusive Methods}

Intrusive gPC methods are non-sampling methods that propagate the
gPC coefficients by solving a single deterministic system of equations
which encapsulates all of the uncertainty information. The stochastic
Galerkin (SGS) method [6] falls under the category of intrusive
UQ methods and is discussed in the context of a Newton's method used
to solve Eq. \hyperlink{eq29}{2.9}. Firstly, we define a gPC approximation of
$\boldsymbol{f}$ as follows. 
\begin{equation}
\boldsymbol{f}\left(\boldsymbol{u}\left(\boldsymbol{\xi}\right),\boldsymbol{v}\left(\boldsymbol{\xi}\right),\boldsymbol{\xi}\right)\approx\boldsymbol{f}^{p}\left(\boldsymbol{\xi}\right)=\sum_{j=0}^{P}\hat{\boldsymbol{f}}^{j}\psi^{j}\left(\boldsymbol{\xi}\right)=\hat{\boldsymbol{F}}\boldsymbol{\psi}\left(\boldsymbol{\xi}\right),
\end{equation}
where 
\begin{equation}
\hat{\boldsymbol{F}}=\hat{\boldsymbol{F}}^{p}=\left[\begin{array}{ccc}
\hat{\boldsymbol{f}}^{0} & \cdots & \hat{\boldsymbol{f}}^{P}\end{array}\right]=\int_{\Xi}\boldsymbol{f}\left(\hat{\boldsymbol{U}}\boldsymbol{\psi}\left(\boldsymbol{\xi}\right),\hat{\boldsymbol{V}}\boldsymbol{\psi}\left(\boldsymbol{\xi}\right),\boldsymbol{\xi}\right)\left(\boldsymbol{\psi}\left(\boldsymbol{\xi}\right)\right)^\mathbf{T}d\mathcal{P}\left(\boldsymbol{\xi}\right).
\end{equation}

Therefore, the deterministic system of equations can be formulated
as follows \hypertarget{eq218}{}
\begin{align}
\hat{\boldsymbol{f}}^{0}\left(\hat{\boldsymbol{u}}^{0},\ldots,\hat{\boldsymbol{u}}^{P},\hat{\boldsymbol{V}}\right) & =\boldsymbol{0},\nonumber \\
 & \vdots\\
\hat{\boldsymbol{f}}^{P}\left(\hat{\boldsymbol{u}}^{0},\ldots,\hat{\boldsymbol{u}}^{P},\hat{\boldsymbol{V}}\right) & =\boldsymbol{0}.\nonumber 
\end{align}

A Newton's method, for instance can be used to solve the Eq. \hyperlink{eq218}{2.18} as follows.\hypertarget{eq219}{}
\begin{equation}
\left[\begin{array}{ccc}
\frac{\partial\boldsymbol{f}^{0}}{\partial\boldsymbol{u}^{0}} & \cdots & \frac{\partial\boldsymbol{f}^{0}}{\partial\boldsymbol{u}^{p}}\\
\vdots &  & \vdots\\
\frac{\partial\boldsymbol{f}^{P}}{\partial\boldsymbol{u}^{0}} & \cdots & \frac{\partial\boldsymbol{f}^{P}}{\partial\boldsymbol{u}^{P}}
\end{array}\right]\left[\begin{array}{c}
\Delta\hat{\boldsymbol{u}}^{0}\\
\vdots\\
\Delta\hat{u}^{P}
\end{array}\right]=-\left[\begin{array}{c}
\hat{\boldsymbol{f}}^{0}\\
\vdots\\
\hat{\boldsymbol{f}}^{P}
\end{array}\right].
\end{equation}

Moreover, we define the gPC approximation of the Jacobian $\boldsymbol{J}=\frac{\partial\boldsymbol{f}}{\partial\boldsymbol{u}}$
in Eq. \hyperlink{eq29}{2.9} as follows. 
\begin{equation}
\boldsymbol{J}\left(\boldsymbol{u}\left(\boldsymbol{\xi}\right),\boldsymbol{v}\left(\boldsymbol{\xi}\right),\boldsymbol{\xi}\right)\approx\boldsymbol{J}^{p}\left(\boldsymbol{\xi}\right)=\sum_{j=0}^{P}\hat{\boldsymbol{J}}^{j}\psi^{j}\left(\boldsymbol{\xi}\right),
\end{equation}
where $\forall0\leq j\leq P$, 
\begin{equation}
\hat{\boldsymbol{J}}^{j}=\int_{\Xi}\boldsymbol{J}\left(\hat{\boldsymbol{U}}\boldsymbol{\psi}\left(\boldsymbol{\xi}\right),\hat{\boldsymbol{V}}\boldsymbol{\psi}\left(\boldsymbol{\xi}\right),\boldsymbol{\xi}\right)\psi^{j}\left(\boldsymbol{\xi}\right)d\mathcal{P}\left(\boldsymbol{\xi}\right).
\end{equation}

Therefore, $\forall0\leq j,k\leq P$, the $\left(j,k\right)$-th block
(of size $n\times n$) in the Jacobian matrix in Eq. \hyperlink{eq219}{2.19} can
be evaluated as follows. 
\begin{equation}
\frac{\partial\boldsymbol{f}^{j}}{\partial\boldsymbol{u}^{k}}=\sum_{l=0}^{P}\hat{\boldsymbol{J}}^{j}c_{jkl}:\forall0\leq l\leq P,c_{jkl}=\int_{\Xi}\psi^{j}\left(\boldsymbol{\xi}\right)\psi^{k}\left(\boldsymbol{\xi}\right)\psi^{l}\left(\boldsymbol{\xi}\right)d\mathcal{P}\left(\boldsymbol{\xi}\right).
\end{equation}

Even with the additional overheads of modifying codes from the deterministic
solver, solving a single deterministic system in lieu of repeated
sampling can be computationally attractive in many instances. However,
intrusive methods based on Galerkin projections in the entire global
polynomial space are impractical for tackling complex multi-physics
systems, where a multidisciplinary developmental strategy is typically
employed and ideally, future updates within a particular module should
not affect the development of other modules.

\section{Module-based hybrid framework for gPC-based UQ}
\hypertarget{sec3}{}
In this section we describe our proposed modular hybrid framework,
where we address this particular issue. Within our proposed module-based
hybrid framework, individual modules (even those that use intrusive
propagation methods) can be developed and managed independently, and
incorporated in a plug-and-play fashion (Figure \hyperlink{fig1}{1}). 

\hypertarget{fig1}{}
\begin{figure}[ph]
    \centering
    \includegraphics[viewport=0bp 0bp 695bp 367bp,clip,scale=0.6]{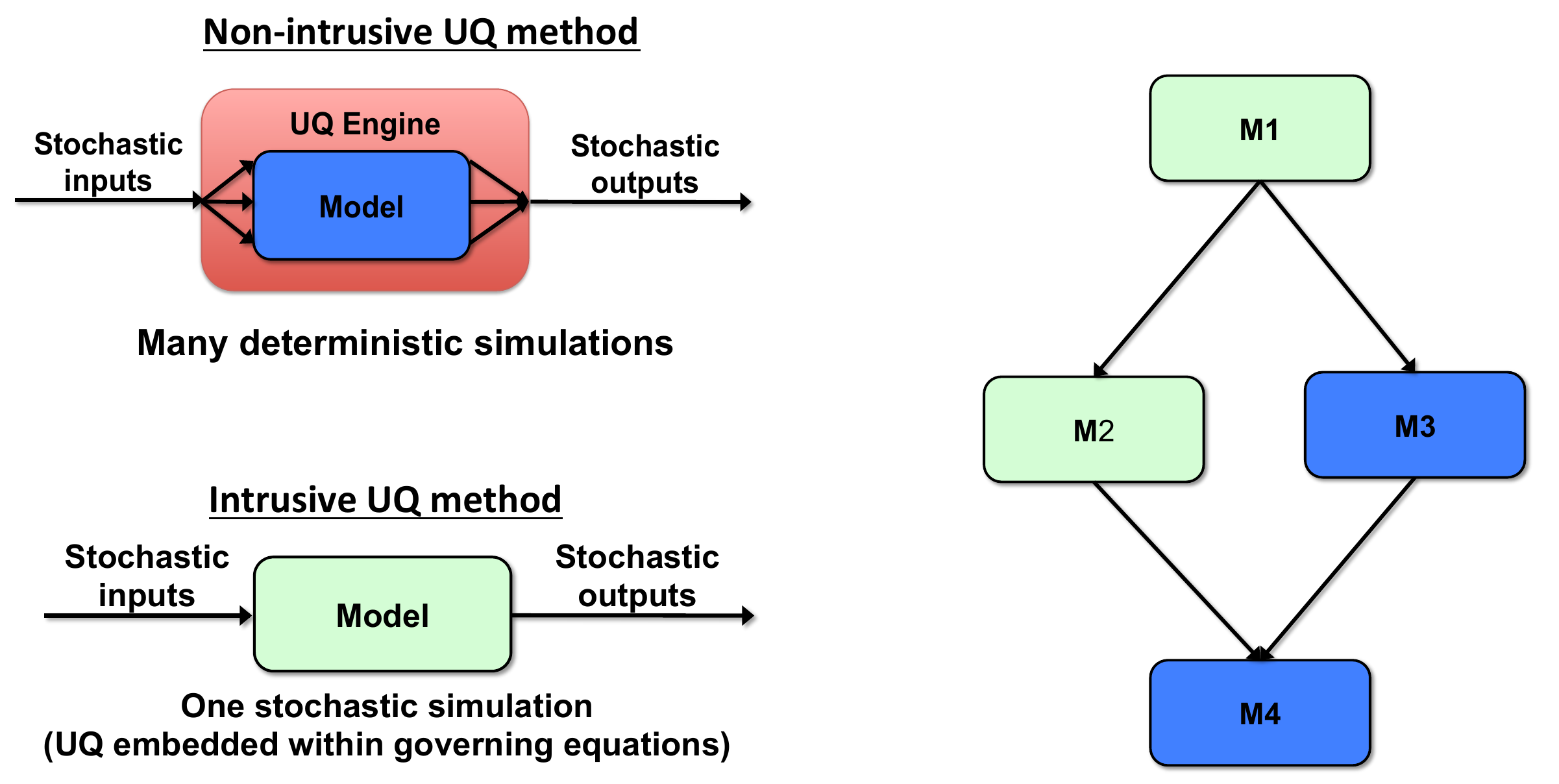}
    \caption{Conceptual illustration of plug-and-play for uncertainty
propagation.}
    \label{fig:fig1}
\end{figure}

Multi-physics models can afford a natural decomposition into subsidiary
single-physics models for which modeling expertise and legacy solvers
may already exist. From practical considerations, the algorithmic
flexibility and modular independence facilitated by this decomposition
can significantly reduce developmental costs and overheads for future
updates. With the added layer of computational complexity entailed
by UQ tasks in predictive simulations, the benefits of modularization
become indispensable. Therefore, we propose a framework which facilitates
the use of the best suited method of propagating local uncertainty
and sensitivity information within each module. To make our discussion
concrete, we consider a generic two-module multi-physics model which
is bidirectionally coupled i.e. Eq. \hyperlink{eq11}{1.1} with $M=2$, and the
corresponding staggered solution method employed. Therefore, at each
iteration $\ell$,
\begin{align}
\boldsymbol{u}_{1}^{\ell+1}\left(\boldsymbol{\xi}_{1},\boldsymbol{\xi}_{2}\right) & =\boldsymbol{m}_{1}\left(\boldsymbol{u}_{1}^{\ell}\left(\boldsymbol{\xi}_{1},\boldsymbol{\xi}_{2}\right),\boldsymbol{u}_{2}^{\ell}\left(\boldsymbol{\xi}_{1},\boldsymbol{\xi}_{2}\right),\boldsymbol{\xi}_{1}\right)\nonumber \\
 & =\boldsymbol{m}_{1}\left(\boldsymbol{y}_{1}^{\ell}\left(\boldsymbol{\xi}_{1},\boldsymbol{\xi}_{2}\right),\boldsymbol{\xi}_{1}\right),\nonumber \\
\boldsymbol{u}_{2}^{\ell+1}\left(\boldsymbol{\xi}_{1},\boldsymbol{\xi}_{2}\right) & =\boldsymbol{m}_{2}\left(\boldsymbol{u}_{1}^{\ell+1}\left(\boldsymbol{\xi}_{1},\boldsymbol{\xi}_{2}\right),\boldsymbol{u}_{2}^{\ell}\left(\boldsymbol{\xi}_{1},\boldsymbol{\xi}_{2}\right),\boldsymbol{\xi}_{2}\right)\nonumber \\
 & =\boldsymbol{m}_{2}\left(\boldsymbol{y}_{2}^{\ell}\left(\boldsymbol{\xi}_{1},\boldsymbol{\xi}_{2}\right),\boldsymbol{\xi}_{2}\right),
\end{align}

where $\boldsymbol{y}_{1}^{\ell}=\left[\boldsymbol{u}_{1}^{\ell};\boldsymbol{u}_{2}^{\ell}\right]$
and $\boldsymbol{y}_{2}^{\ell}=\left[\boldsymbol{u}_{1}^{\ell+1};\boldsymbol{u}_{2}^{\ell}\right]$.
Here, we have used the Gauss-Seidel iterative method, which usually
performs better than the standard Jacobi method. The dimensions of
the solution fields, operators and stochastic parameters are assumed
to be consistent with those defined with respect to the $M-$module
system (1). Since their respective input and output quantities that
depend on both $\boldsymbol{\xi}_{1}$ and $ $$\boldsymbol{\xi}_{2}$,
each module would need to deal with uncertainties that are external
to its local parameter space. In general, we can represent the stochastic
module operators as follows. $\forall i\in\left\{ 1,2\right\} ,$
and iteration $\ell$, 
\begin{equation}
\hat{\boldsymbol{U}}_{i}^{\ell+1}=\boldsymbol{M}_{i}\left(\hat{\boldsymbol{Y}}_{i}^{\ell}\right),
\end{equation}
where $\hat{\boldsymbol{Y}}_{1}^{\ell}=\left[\hat{\boldsymbol{U}}_{1}^{\ell};\hat{\boldsymbol{U}}_{2}^{\ell}\right]$
and $\hat{\boldsymbol{Y}}_{2}^{\ell}=\left[\hat{\boldsymbol{U}}_{1}^{\ell+1};\hat{\boldsymbol{U}}_{2}^{\ell}\right]$.

In a monolithic intrusive propagation framework, any changes made
in characterizing the probability spaces of $\boldsymbol{\xi}_{1}$
or $\boldsymbol{\xi}_{2}$ would need to be reflected in both $\boldsymbol{M}_{1}$
and $\boldsymbol{M}_{2}$. In our proposed framework, our algorithmic
goal is to make each module ``self-inclusive'' in its implementation
and flexible in its methodology for propagating the local stochastic
information. Thereby, we would retain module independence and any
changes made to the set of local uncertainties in a module would not
affect the development of other modules. Towards this end, we need
to derive the necessary operators which transform between global and
modular representations of stochastic information. In general, we
can represent these transformations as follows.
\hypertarget{eq33}{}
\begin{align}
\hat{\boldsymbol{U}}_{i}^{\ell+1} & =\sum_{j=1}^{N_{1}}\Phi_{ij}^{-1}\circ\tilde{\boldsymbol{M}}_{ij}\left(\Phi_{ij}\circ\hat{\boldsymbol{Y}}_{i}^{\ell}\right),
\end{align}
where $\Phi_{ij}$ denotes the $restriction$ map in module $i$ to
subproblem $j$ and $\Phi_{ij}^{-1}$ denotes the reverse $prolongation$
map. Moreover, $\tilde{\boldsymbol{M}}_{ij}$ represents the stochastic
module operator based on $\boldsymbol{m}_{i}$ which depends on $\boldsymbol{\xi}_{i}$
or $\Xi_{i}$, depending on the type of uncertainty propagation
method employed.

For linear multi-physics models, a decomposition property of linear
modules can be exploited such that the computation of the gPC coefficients
at the next iteration can be split into independent subproblems of
smaller size (corresponding to different external indices for the
external polynomial basis functions). Besides the truncation error
imposed by the gPC approximation, this decomposition does not suffer
from any additional loss of stochastic information. A proof of this
property has been provided in the previous work [1], but limited
to the case when the input parameters in each module are scalars.
Lemma \hyperlink{lemA3}{A3} and Lemma \hyperlink{lemA4}{A4} in Appendix A prove the general case when the
input parameters are vectors, in the context of non-intrusive and
intrusive projection methods respectively.

We now consider a general nonlinear model setup for intrusive spectral
projection, for which the corresponding restriction and prolongation
maps will be defined. In each module, the local stochastic information
can be represented using a $modular$ gPC approximation of the input/output
data.

\subsection{Modular gPC approximation}

For a given order $p$, let $\boldsymbol{\psi}\equiv\boldsymbol{\psi}^{p}$
denote as the global polynomial basis vector with $\boldsymbol{\psi}_{1}\equiv\boldsymbol{\psi}_{1}^{p}$
and $\boldsymbol{\psi}_{2}\equiv\boldsymbol{\psi}_{2}^{p}$ denoting
the modular polynomial basis vectors. Therefore, we can relate the
global and modular gPC approximations as follows. $\forall i\in\left\{ 1,2\right\} ,$
\begin{align}
\boldsymbol{u}_{i}^{p}\left(\boldsymbol{\xi}_{1},\boldsymbol{\xi}_{2}\right) & =\sum_{\left|\boldsymbol{j}\right|=0}^{p}\hat{\boldsymbol{u}}_{i}^{\boldsymbol{j}_{1}\boldsymbol{j}_{2}}\psi_{1}^{\boldsymbol{j}_{1}}\left(\boldsymbol{\xi}_{1}\right)\psi_{2}^{\boldsymbol{j}_{2}}\left(\boldsymbol{\xi}_{2}\right)=\hat{\boldsymbol{U}}_{i}\boldsymbol{\psi}\left(\boldsymbol{\xi}_{1},\boldsymbol{\xi}_{2}\right)\nonumber \\
 & =\sum_{\left|\boldsymbol{j}\right|=0}^{p}\tilde{\boldsymbol{u}}_{i,1}^{\boldsymbol{j}}\left(\boldsymbol{\xi}_{2}\right)\psi_{1}^{\boldsymbol{j}}\left(\boldsymbol{\xi}_{1}\right)=\tilde{\boldsymbol{U}}_{i,1}\left(\boldsymbol{\xi}_{2}\right)\boldsymbol{\psi}_{1}\left(\boldsymbol{\xi}_{1}\right)\nonumber \\
 & =\sum_{\left|\boldsymbol{j}\right|=0}^{p}\tilde{\boldsymbol{u}}_{i,2}^{\boldsymbol{j}}\left(\boldsymbol{\xi}_{1}\right)\psi_{2}^{\boldsymbol{j}}\left(\boldsymbol{\xi}_{2}\right)=\tilde{\boldsymbol{U}}_{i,2}\left(\boldsymbol{\xi}_{1}\right)\boldsymbol{\psi}_{2}\left(\boldsymbol{\xi}_{2}\right),
\end{align}
where $\tilde{\boldsymbol{U}}_{i,1}$ and $\tilde{\boldsymbol{U}}_{i,2}$
are the modular gPC coefficient matrices in module 1 and module 2
respectively. Subsequently, the restriction and prolongation maps
are defined to transform between global and modular gPC matrices. 

Since implementation of the intrusive stochastic modules is only based
on the local stochastic information contained in the inputs and output
data, the restriction map transforms the global gPC coefficient matrix
to the module gPC coefficient matrix at various sampling points in
the external stochastic parameter space. Let $P+1={s+p \choose p}$
denote the total number of global basis polynomials and $P_{i}+1={s_{i}+p \choose p}:i\in\left\{ 1,2\right\} $
denote the total number of modular basis polynomials in module $i$. 

Considering module 1, for instance, we define the set of samples $\left\{ \boldsymbol{\xi}_{2}^{\left(j\right)}\right\} _{j=1}^{Q_{2}}$.
Subsequently, we can define the restriction map as follows. $\forall i\in\left\{ 1,2\right\} ,1\leq j\leq Q_{2},$\hypertarget{eq35}{}
\begin{equation}
\tilde{\boldsymbol{U}}_{i,1}\left(\boldsymbol{\xi}_{2}^{\left(j\right)}\right)=\Phi_{1j}\circ\hat{\boldsymbol{U}}_{i}=\hat{\boldsymbol{U}}_{i}\boldsymbol{P}_{1}\boldsymbol{\Pi}_{1}\left(\boldsymbol{\xi}_{2}^{\left(j\right)}\right),
\end{equation}
where $\boldsymbol{\Pi}_{1}:\forall\boldsymbol{\xi}_{2}\in\Xi_{2},$
\begin{equation}
\boldsymbol{\Pi}_{1}\left(\boldsymbol{\xi}_{2}\right)=\left[\begin{array}{ccc}
\boldsymbol{\psi}_{2}^{p-\left|\boldsymbol{j}_{0}\right|}\left(\boldsymbol{\xi}_{2}\right)\\
 & \ddots\\
 &  & \boldsymbol{\psi}_{2}^{p-\left|\boldsymbol{j}_{P_{1}}\right|}\left(\boldsymbol{\xi}_{2}\right)
\end{array}\right]
\end{equation}
is a $\left(P+1\right)\times\left(P_{1}+1\right)$ sparse transformation
matrix with at most $P$ non-zero entries. Moreover, $\boldsymbol{P}_{1}$
is the corresponding permutation matrix and $\forall0\leq k\leq P_{1}$,
$\boldsymbol{j}_{k}\in\mathbb{N}_{0}^{s_{1}}:0\leq\left|\boldsymbol{j}_{k}\right|\leq p$
and $\left|\boldsymbol{j}_{0}\right|\leq\cdots\leq\left|\boldsymbol{j}_{P_{1}}\right|$. 

For any $\boldsymbol{u}:\Xi\rightarrow\mathbb{R}^{n}$, the corresponding
global gPC coefficient matrix can be approximated as follows.\hypertarget{eq37}{}
\begin{equation}
\hat{\boldsymbol{U}}\approx\sum_{i=1}^{Q_{2}}\Phi_{1j}^{-1}\circ\tilde{\boldsymbol{U}}_{1}\left(\boldsymbol{\xi}_{2}^{\left(j\right)}\right)\ \mathrm{or}\ \hat{\boldsymbol{U}}\approx\sum_{i=1}^{Q_{1}}\Phi_{2j}^{-1}\circ\tilde{\boldsymbol{U}}_{2}\left(\boldsymbol{\xi}_{1}^{\left(j\right)}\right)
\end{equation}
where the prolongation map $\Phi_{ij}^{-1}$ can be defined using
either a least-squares regression the pseudospectral approach, as
provided Theorem \hyperlink{thm1}{1} and Theorem \hyperlink{thm2}{2} respectively.

\hypertarget{thm1}{}
\subsubsection*{Theorem 1:}

Given a sufficiently large number of samples of the modular gPC coefficient
matrices $\left\{ \tilde{\boldsymbol{U}}_{1}\left(\boldsymbol{\xi}_{2}^{\left(j\right)}\right)\right\} _{j=1}^{Q_{2}}$
of a function $\boldsymbol{u}:\Xi\rightarrow\mathbb{R}^{n}$, a
least-square recovery of the global gPC coefficient matrix can be
obtained, according to Eq. \hyperlink{eq37}{3.7}, using the prolongation map $\Phi_{1j}^{-1}:$
$\forall1\leq j\leq Q_{2},$\hypertarget{eq38}{}
\begin{equation}
\Phi_{1j}^{-1}\circ\tilde{\boldsymbol{U}}_{1}\left(\boldsymbol{\xi}_{2}^{\left(j\right)}\right)=\tilde{\boldsymbol{U}}_{1}\left(\boldsymbol{\xi}_{2}^{\left(j\right)}\right)\boldsymbol{\Pi}_{1}\left(\boldsymbol{\xi}_{2}^{\left(j\right)}\right)^\mathbf{T}\boldsymbol{V}_{1}^{-1}\boldsymbol{P}_{1}^\mathbf{T},
\end{equation}
where
\begin{equation}
\boldsymbol{V}_{1}=\sum_{j=1}^{Q_{2}}\boldsymbol{\Pi}_{1}\left(\boldsymbol{\xi}_{2}^{\left(j\right)}\right)\boldsymbol{\Pi}_{1}\left(\boldsymbol{\xi}_{2}^{\left(j\right)}\right)^\mathbf{T}.
\end{equation}

\subsubsection*{Proof:}

A least-squares recovery on the global gPC coefficient matrix can
be formulated as the following minimization problem.
\begin{align}
\hat{\boldsymbol{U}} & =\arg\min_{\hat{\boldsymbol{Y}}\in\mathbb{R}^{n\times\left(P+1\right)}}\left\Vert \left[\begin{array}{c}
\tilde{\boldsymbol{U}}_{1}\left(\boldsymbol{\xi}_{2}^{\left(1\right)}\right)-\hat{\boldsymbol{Y}}\boldsymbol{P}_{1}\boldsymbol{\Pi}_{1}\left(\boldsymbol{\xi}_{2}^{\left(1\right)}\right)\\
\vdots\\
\tilde{\boldsymbol{U}}_{1}\left(\boldsymbol{\xi}_{2}^{\left(Q\right)}\right)-\hat{\boldsymbol{Y}}\boldsymbol{P}_{1}\boldsymbol{\Pi}_{1}\left(\boldsymbol{\xi}_{2}^{\left(Q\right)}\right)
\end{array}\right]\right\Vert _{F}.
\end{align}
The derivation in Lemma \hyperlink{lemA1}{A1} can be used to prove Eq.\hyperlink{eq38}{3.8} as follows.\hypertarget{eq311}{}
\begin{align}
\hat{\boldsymbol{U}} & =\left[\begin{array}{ccc}
\tilde{\boldsymbol{U}}_{1}\left(\boldsymbol{\xi}_{2}^{\left(1\right)}\right) & \cdots & \tilde{\boldsymbol{U}}_{1}\left(\boldsymbol{\xi}_{2}^{\left(Q\right)}\right)\end{array}\right]\left[\begin{array}{c}
\boldsymbol{\Pi}_{1}\left(\boldsymbol{\xi}_{2}^{\left(1\right)}\right)\\
\vdots\\
\boldsymbol{\Pi}_{1}\left(\boldsymbol{\xi}_{2}^{\left(Q\right)}\right)
\end{array}\right]^\mathbf{T}\nonumber \\
 & \times\left(\left[\begin{array}{ccc}
\boldsymbol{\Pi}_{1}\left(\boldsymbol{\xi}_{2}^{\left(1\right)}\right) & \cdots & \boldsymbol{\Pi}_{1}\left(\boldsymbol{\xi}_{2}^{\left(Q\right)}\right)\end{array}\right]\left[\begin{array}{c}
\boldsymbol{\Pi}_{1}\left(\boldsymbol{\xi}_{2}^{\left(1\right)}\right)\\
\vdots\\
\boldsymbol{\Pi}_{1}\left(\boldsymbol{\xi}_{2}^{\left(Q\right)}\right)
\end{array}\right]^\mathbf{T}\right)^{-1}\boldsymbol{P}_{1}^\mathbf{T}\nonumber \\
 & =\left[\begin{array}{ccc}
\tilde{\boldsymbol{U}}_{1}\left(\boldsymbol{\xi}_{2}^{\left(1\right)}\right) & \cdots & \tilde{\boldsymbol{U}}_{1}\left(\boldsymbol{\xi}_{2}^{\left(Q\right)}\right)\end{array}\right]\left[\begin{array}{c}
\boldsymbol{\Pi}_{1}\left(\boldsymbol{\xi}_{2}^{\left(1\right)}\right)\\
\vdots\\
\boldsymbol{\Pi}_{1}\left(\boldsymbol{\xi}_{2}^{\left(Q\right)}\right)
\end{array}\right]^\mathbf{T}\boldsymbol{V}_{1}^{-1}\boldsymbol{P}_{1}^\mathbf{T}\nonumber \\
 & =\sum_{j=1}^{Q_{2}}\tilde{\boldsymbol{U}}_{1}\left(\boldsymbol{\xi}_{2}^{\left(j\right)}\right)\boldsymbol{\Pi}_{1}\left(\boldsymbol{\xi}_{2}^{\left(j\right)}\right)^\mathbf{T}\boldsymbol{V}_{1}^{-1}\boldsymbol{P}_{1}^\mathbf{T}.
\end{align}

Therefore, combining Eq. \hyperlink{eq37}{3.7} and Eq. \hyperlink{eq311}{3.11} yields Eq. \hyperlink{eq38}{3.8}. $\square$

\hypertarget{thm2}{}
\subsubsection*{Theorem 2:}

Given a quadrature rule $\left\{ \left(\boldsymbol{\xi}_{2}^{\left(j\right)},w_{2}^{\left(j\right)}\right)\right\} _{j=1}^{Q_{2}}$,
which can exactly integrate polynomials with total degree $\leq2p$,
and the corresponding samples of the modular gPC coefficient matrices
$\left\{ \tilde{\boldsymbol{U}}_{1}\left(\boldsymbol{\xi}_{2}^{\left(j\right)}\right)\right\} _{j=1}^{Q_{2}}$
of a function $\boldsymbol{u}:\Xi\rightarrow\mathbb{R}^{n}$, a
pseudospectral recovery of the global gPC coefficient matrix can be
obtained, according to Eq. 3.7, using the prolongation map $\Phi_{1j}^{-1}:$ $\forall1\leq j\leq Q_{2},$\hypertarget{eq312}{}
\begin{equation}
\Phi_{1j}^{-1}\circ\tilde{\boldsymbol{U}}_{1}\left(\boldsymbol{\xi}_{2}^{\left(j\right)}\right)=w_{2}^{\left(j\right)}\tilde{\boldsymbol{U}}_{1}\left(\boldsymbol{\xi}_{2}^{\left(j\right)}\right)\boldsymbol{\Pi}_{1}\left(\boldsymbol{\xi}_{2}^{\left(j\right)}\right)^\mathbf{T}\boldsymbol{P}_{1}^\mathbf{T}.
\end{equation}

\subsubsection*{Proof:}

Since $\boldsymbol{\Pi}_{1}$ contains polynomials of total degree
less than or equal to $p$, we have
\begin{equation}
\int_{\mathbb{R}^{s_{2}}}\boldsymbol{\Pi}_{1}\left(\boldsymbol{\xi}_{2}\right)\boldsymbol{\Pi}_{1}\left(\boldsymbol{\xi}_{2}\right)^\mathbf{T}d\mathcal{P}\left(\boldsymbol{\xi}_{2}\right)=\sum_{j=1}^{Q}w_{2}^{\left(j\right)}\boldsymbol{\Pi}_{1}\left(\boldsymbol{\xi}_{2}^{\left(j\right)}\right)\boldsymbol{\Pi}_{1}\left(\boldsymbol{\xi}_{2}^{\left(j\right)}\right)^\mathbf{T}=\boldsymbol{I}_{P+1}.
\end{equation}

Therefore, we have\hypertarget{eq314}{}
\begin{align}
\hat{\boldsymbol{U}} & =\hat{\boldsymbol{U}}\boldsymbol{P}_{1}\left(\int_{\mathbb{R}^{s_{2}}}\boldsymbol{\Pi}_{1}\left(\boldsymbol{\xi}_{2}\right)\boldsymbol{\Pi}_{1}\left(\boldsymbol{\xi}_{2}\right)^\mathbf{T}d\mathcal{P}\left(\boldsymbol{\xi}_{2}\right)\right)\boldsymbol{P}_{1}^\mathbf{T}\nonumber \\
 & =\left(\int_{\mathbb{R}^{s_{2}}}\hat{\boldsymbol{U}}\boldsymbol{P}_{1}\boldsymbol{\Pi}_{1}\left(\boldsymbol{\xi}_{2}\right)\boldsymbol{\Pi}_{1}\left(\boldsymbol{\xi}_{2}\right)^\mathbf{T}d\mathcal{P}\left(\boldsymbol{\xi}_{2}\right)\right)\boldsymbol{P}_{1}^\mathbf{T}\nonumber \\
 & =\left(\mbox{\ensuremath{\int_{\mathbb{R}^{s_{2}}}\tilde{\boldsymbol{U}}_{1}\left(\boldsymbol{\xi}_{2}\right)\boldsymbol{\Pi}_{1}\left(\boldsymbol{\xi}_{2}\right)^\mathbf{T}}d\ensuremath{\mathcal{P}\left(\boldsymbol{\xi}_{2}\right)}}\right)\boldsymbol{P}_{1}^\mathbf{T}\nonumber \\
 & =\sum_{j=1}^{Q_{2}}w_{2}^{\left(j\right)}\tilde{\boldsymbol{U}}_{1}\left(\boldsymbol{\xi}_{2}^{\left(j\right)}\right)\boldsymbol{\Pi}_{1}\left(\boldsymbol{\xi}_{2}^{\left(j\right)}\right)^\mathbf{T}\boldsymbol{P}_{1}^\mathbf{T}.
\end{align}
Therefore, combining Eq. \hyperlink{eq37}{3.7} and Eq. \hyperlink{eq314}{3.14} yields Eq. \hyperlink{eq312}{3.12}. $\square$

Similarly, we can define the restriction and prolongation maps corresponding
to module 2. These maps provide the basic components of our proposed
module-based hybrid framework and their specific definition depends on
the problem structure and uncertainty propagation method employed
in each module. In the context of Eq. \hyperlink{eq33}{3.3}. we will now summarize
the these definitions for the uncertainty propagation methods discussed
in section 2.

\subsection{Restriction and prolongation map definitions}

For linear problems, we have $N_{1}=P_{2}+1$, $N_{2}=P_{1}+1$. $\forall i\in\left\{ 1,2\right\} $,
let $\mathcal{J}_{i}\equiv\left\{ \boldsymbol{j}_{j}\in\mathbb{N}_{0}^{s_{i}}:0\leq\left|\boldsymbol{j}_{j}\right|\leq p\right\} _{j=1}^{P_{i}+1}$
denote the internal index set, $\mathcal{K}_{i}\equiv\left\{ \boldsymbol{k}_{j}\in\mathbb{N}_{0}^{s-s_{i}}:0\leq\left|\boldsymbol{k}_{j}\right|\leq p\right\} _{j=1}^{N_{i}}$
denote the external index set and $\mathcal{I}_{i}\equiv\left\{ \boldsymbol{j}_{j}\boldsymbol{k}_{j}\right\} _{j=1}^{P+1}$
denote the global index set. Subsequently, $\forall i\in\left\{ 1,2\right\} $,$1\leq j\leq N_{i},$and iteration $\ell$, we have
\begin{align}
\Phi_{ij}\circ\hat{\boldsymbol{Y}}_{i}^{\ell} & =\hat{\boldsymbol{Y}}_{i}^{\ell}\boldsymbol{P}_{i}\boldsymbol{E}_{j},\nonumber \\
\Phi_{ij}^{-1}\circ\tilde{\boldsymbol{M}}_{ij} & =\tilde{\boldsymbol{M}}_{ij}\boldsymbol{E}_{j}^\mathbf{T}\boldsymbol{P}_{i}^\mathbf{T},
\end{align}
where the columns in $\boldsymbol{E}_{j}$ are a subset of the columns
(unit vectors) in $\boldsymbol{I}_{P+1}$ such that
\begin{equation}
\boldsymbol{E}_{j}=\left[\begin{array}{ccc}
\cdots & \boldsymbol{e}_{k} & \cdots\end{array}\right]:\boldsymbol{j}_{k}\boldsymbol{k}_{j}\in\mathcal{I}_{i}.
\end{equation}

Moreover, we define $\hat{\boldsymbol{Y}}_{i,j}^{\ell}=\Phi_{ij}\circ\hat{\boldsymbol{Y}}_{i}^{\ell}$,
$\boldsymbol{\psi}_{i,j}=\boldsymbol{\psi}_{i}^{p-\left|\boldsymbol{k}_{j}\right|}$,
$\boldsymbol{\Psi}_{i,j}=\boldsymbol{\Psi}_{i}^{p-\left|\boldsymbol{k}_{j}\right|}$, $\tilde{\boldsymbol{\psi}}_{i,j}=\tilde{\boldsymbol{\psi}}_{i}^{p-\left|\boldsymbol{k}_{j}\right|}$,
$\tilde{\boldsymbol{\Psi}}_{i,j}=\tilde{\boldsymbol{\Psi}}_{i}^{p-\left|\boldsymbol{k}_{j}\right|}$ and $\tilde{\boldsymbol{m}}_{i}$
as the modified $i$-th module with its outputs $\left[\begin{array}{cc}
\boldsymbol{u}_{i}^{\ell+1} & \frac{\partial\boldsymbol{u}_{i}^{\ell+1}}{\partial\boldsymbol{\xi}_{i}}\end{array}\right]$. Subsequently, we can define $\tilde{\boldsymbol{M}}_{ij}$, based
on the uncertainty propagation method used in module $i$, as follows.
\begin{itemize}
\item[I.] Non-intrusive regression: We have
\begin{align}
\tilde{\boldsymbol{M}}_{ij}\left(\hat{\boldsymbol{Y}}_{i,j}^{\ell}\right) & =\left(\sum_{l=1}^{Q_{i,j}}\boldsymbol{m}_{i}\left(\hat{\boldsymbol{Y}}_{i,j}^{\ell}\boldsymbol{\psi}_{i,j}\left(\boldsymbol{\xi}_{i}^{\left(l\right)}\right)\right)\boldsymbol{\psi}_{i,j}\left(\boldsymbol{\xi}_{i}^{\left(l\right)}\right)^\mathbf{T}\right)\left(\boldsymbol{\Psi}_{i,j}\boldsymbol{\Psi}_{i,j}^\mathbf{T}\right)^{-1}
\end{align}

\item[II.] Non-intrusive projection: We have 
\begin{equation}
\tilde{\boldsymbol{M}}_{ij}\left(\hat{\boldsymbol{Y}}_{i,j}^{\ell}\right)=\sum_{l=1}^{Q_{i,j}}w_{i}^{\left(l\right)}\boldsymbol{m}_{i}\left(\hat{\boldsymbol{Y}}_{i,j}^{\ell}\boldsymbol{\psi}_{i,j}\left(\boldsymbol{\xi}_{i}^{\left(l\right)}\right)\right)\boldsymbol{\psi}_{i,j}\left(\boldsymbol{\xi}_{i}^{\left(l\right)}\right)^\mathbf{T}.
\end{equation}
\item[III.] Semi-intrusive regression: We have
\begin{align}
\tilde{\boldsymbol{M}}_{ij}\left(\hat{\boldsymbol{Y}}_{i,j}^{\ell}\right) & =\left(\sum_{l=1}^{Q_{i,j}}\tilde{\boldsymbol{m}}_{i}\left(\hat{\boldsymbol{Y}}_{i,j}^{\ell}\tilde{\boldsymbol{\psi}}_{i,j}\left(\boldsymbol{\xi}_{i}^{\left(l\right)}\right)\right)\tilde{\boldsymbol{\psi}}_{i,j}\left(\boldsymbol{\xi}_{i}^{\left(l\right)}\right){}^\mathbf{T}\right)\left(\tilde{\boldsymbol{\Psi}}_{i.j}\tilde{\boldsymbol{\Psi}}_{i.j}^\mathbf{T}\right)^{-1}.
\end{align}

\item[IV.] Intrusive projection: From Lemma \hyperlink{lemA4}{A4}, we have
\begin{align}
 & \left(\int_{\Xi_{i}}\left(\boldsymbol{\psi}_{i,j}\left(\boldsymbol{\xi}_{i}\right)\boldsymbol{\psi}_{i,j}\left(\boldsymbol{\xi}_{i}\right)^\mathbf{T}\right)\otimes\boldsymbol{A}_{i}\left(\boldsymbol{\xi}_{i}\right)d\mathcal{P}_{i}\left(\boldsymbol{\xi}_{i}\right)\right)vec\left(\tilde{\boldsymbol{M}}_{ij}\left(\hat{\boldsymbol{Y}}_{i,j}^{\ell}\right)\right)\nonumber \\
= & \int_{\Xi_{i}}\boldsymbol{\psi}_{i,j}\left(\boldsymbol{\xi}_{i}\right)\otimes\boldsymbol{b}_{i}\left(\hat{\boldsymbol{Y}}_{i,j}^{\ell}\boldsymbol{\psi}_{i}^{j}\left(\boldsymbol{\xi}_{i}\right),\boldsymbol{\xi}_{i}\right)d\mathcal{P}_{i}\left(\boldsymbol{\xi}_{i}\right).
\end{align}

\end{itemize}
As opposed to linear problems, the definition of restriction and prolongation
maps in nonlinear problems would depend on the uncertainty propagation
method used in the respective modules. 
\begin{itemize}
\item[I.] Non-intrusive regression: We define $N_{i}=Q$ and 
\begin{align}
\Phi_{ij}\circ\hat{\boldsymbol{Y}}_{i}^{\ell} & =\hat{\boldsymbol{Y}}_{i}^{\ell}\boldsymbol{P}_{i}\boldsymbol{\psi}\left(\boldsymbol{\xi}^{\left(j\right)}\right),\nonumber \\
\tilde{\boldsymbol{M}}_{ij}\left(\Phi_{ij}\circ\hat{\boldsymbol{Y}}_{i}^{\ell}\right) & =\boldsymbol{m}_{i}\left(\Phi_{ij}\circ\hat{\boldsymbol{Y}}_{i}^{\ell},\boldsymbol{\xi}_{i}^{\left(j\right)}\right),\nonumber \\
\Phi_{ij}^{-1}\circ\tilde{\boldsymbol{M}}_{ij} & =\tilde{\boldsymbol{M}}_{ij}\boldsymbol{\psi}\left(\boldsymbol{\xi}^{\left(j\right)}\right)^\mathbf{T}\left(\boldsymbol{\Psi}\boldsymbol{\Psi}^\mathbf{T}\right)^{-1}\boldsymbol{P}_{i}^\mathbf{T}.
\end{align}

\item[II.] Non-intrusive projection: We define $N_{i}=Q$ and 
\begin{align}
\Phi_{ij}\circ\hat{\boldsymbol{Y}}_{i}^{\ell} & =\hat{\boldsymbol{Y}}_{i}^{\ell}\boldsymbol{P}_{i}\boldsymbol{\psi}\left(\boldsymbol{\xi}^{\left(j\right)}\right),\nonumber \\
\tilde{\boldsymbol{M}}_{ij}\left(\Phi_{ij}\circ\hat{\boldsymbol{Y}}_{i}^{\ell}\right) & =\boldsymbol{m}_{i}\left(\Phi_{ij}\circ\hat{\boldsymbol{Y}}_{i}^{\ell},\boldsymbol{\xi}_{i}^{\left(j\right)}\right),\nonumber \\
\Phi_{ij}^{-1}\circ\tilde{\boldsymbol{M}}_{ij} & =w^{\left(j\right)}\tilde{\boldsymbol{M}}_{ij}\boldsymbol{\psi}\left(\boldsymbol{\xi}^{\left(j\right)}\right)^\mathbf{T}\boldsymbol{P}_{i}^\mathbf{T}.
\end{align}

\item[III.] Semi-intrusive regression: We define $N_{i}=Q_{i}$ and 
\begin{align}
\Phi_{ij}\circ\hat{\boldsymbol{Y}}_{i}^{\ell} & =\hat{\boldsymbol{Y}}_{i}^{\ell}\boldsymbol{P}_{i}\tilde{\boldsymbol{\psi}}_{i}\left(\boldsymbol{\xi}^{\left(j\right)}\right),\nonumber \\
\tilde{\boldsymbol{M}}_{ij}\left(\Phi_{ij}\circ\hat{\boldsymbol{Y}}_{i}^{\ell}\right) & =\tilde{\boldsymbol{m}}_{i}\left(\Phi_{ij}\circ\hat{\boldsymbol{Y}}_{i}^{\ell},\boldsymbol{\xi}_{i}^{\left(j\right)}\right),\nonumber \\
\Phi_{ij}^{-1}\circ\tilde{\boldsymbol{M}}_{ij} & =\tilde{\boldsymbol{M}}_{ij}\tilde{\boldsymbol{\psi}}_{i}\left(\boldsymbol{\xi}^{\left(j\right)}\right)^\mathbf{T}\left(\tilde{\boldsymbol{\Psi}}_i\tilde{\boldsymbol{\Psi}}_i^\mathbf{T}\right)^{-1}\boldsymbol{P}_{i}^\mathbf{T}.
\end{align}

\item[IV.] Intrusive projection: We define $N_{1}=Q_{2}$ and $N_{2}=Q_{1}$,
the restriction map $\Phi_{ij}$ using Eq. \hyperlink{eq35}{3.5}, the prolongation
map $\Phi_{ij}^{-1}$ using either Eq. \hyperlink{eq38}{3.8} or Eq. \hyperlink{eq312}{3.12},
and $\tilde{\boldsymbol{M}}_{ij}:$
\end{itemize}
\begin{align}
 & \left(\int_{\Xi_{i}}\left(\boldsymbol{\psi}_{i}\left(\boldsymbol{\xi}_{i}\right)\boldsymbol{\psi}_{i}\left(\boldsymbol{\xi}_{i}\right)^\mathbf{T}\right)\otimes\frac{\partial\boldsymbol{f}_{i}}{\partial\boldsymbol{u}_{i}}\left(\Phi_{ij}\circ\hat{\boldsymbol{Y}}_{i}^{\ell}\boldsymbol{\psi}_{i}\left(\boldsymbol{\xi}_{i}\right),\boldsymbol{\xi}_{i}\right)d\mathcal{P}_{i}\left(\boldsymbol{\xi}_{i}\right)\right)\nonumber \\
 & \times vec\left(\tilde{\boldsymbol{M}}_{ij}\left(\Phi_{ij}\circ\hat{\boldsymbol{Y}}_{i}^{\ell}\right)-\Phi_{ij}\circ\hat{\boldsymbol{U}}_{i}^{\ell}\right)\nonumber \\
 & =-\int_{\Xi_{i}}\boldsymbol{\psi}_{i}\left(\boldsymbol{\xi}_{i}\right)\otimes\boldsymbol{f}_{i}\left(\Phi_{ij}\circ\hat{\boldsymbol{Y}}_{i}^{\ell}\boldsymbol{\psi}_{i}\left(\boldsymbol{\xi}_{i}\right),\boldsymbol{\xi}_{i}\right)d\mathcal{P}_{i}\left(\boldsymbol{\xi}_{i}\right).
\end{align}

\hypertarget{sec4}{}
\section{Numerical example}

In this section, we demonstrate an implementation of our proposed
framework using a thermally driven cavity flow problem as a multi-physics
simulation problem, with uncertain boundary conditions and fluid properties.
\hypertarget{fig2}{}
\begin{figure}[ph]
    \centering
    \includegraphics[bb=125bp 500bp 475bp 720bp,scale=0.65]{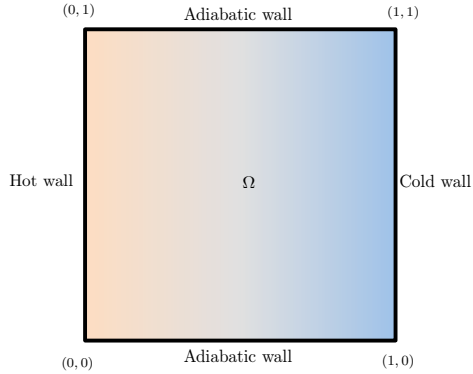}
    \caption{Computational domain for the thermally-driven cavity flow
problem.}
    \label{fig:fig2}
\end{figure}

\subsection{Model setup}

Firstly, we consider a 2D square cavity $\Omega=\left(0,1\right)_{r_{1}}\times\left(0,1\right)_{r_{2}}$
in which the non-dimensional fluid velocity \textbf{$\boldsymbol{u}=\left[\begin{array}{cc}
u_{1} & u_{2}\end{array}\right]^\mathbf{T}$} , pressure $p$ , and temperature $T$ are governed by the incompressible
Boussineseq equations [\hyperlink{ref34}{34}]: \hypertarget{eq41}{}
\begin{align}
\boldsymbol{\nabla}^\mathbf{T}\boldsymbol{u}\left(\boldsymbol{x},\boldsymbol{\xi}\right) & =0,\nonumber \\
\left(\boldsymbol{u}\left(\boldsymbol{x},\boldsymbol{\xi}\right)^\mathbf{T}\boldsymbol{\nabla}\right)\boldsymbol{u}\left(\boldsymbol{x},\boldsymbol{\xi}\right)+\boldsymbol{\nabla}p\left(\boldsymbol{x},\boldsymbol{\xi}\right)\nonumber \\
-\mathrm{Pr}\boldsymbol{\nabla}^\mathbf{T}\boldsymbol{\nabla}\boldsymbol{u}\left(\boldsymbol{x},\boldsymbol{\xi}\right)-\mathrm{Pr}\mathrm{Ra}\left(\boldsymbol{x},\boldsymbol{\xi}_{1}\right)T\left(\boldsymbol{x},\boldsymbol{\xi}\right)\boldsymbol{e}_{2} & =\boldsymbol{0},\nonumber \\
\left(\boldsymbol{u}\left(\boldsymbol{x},\boldsymbol{\xi}\right)^\mathbf{T}\boldsymbol{\nabla}\right)T\left(\boldsymbol{x},\boldsymbol{\xi}\right)-\boldsymbol{\nabla}^\mathbf{T}\boldsymbol{\nabla}T\left(\boldsymbol{x},\boldsymbol{\xi}\right) & =0, & \boldsymbol{x}\in\Omega,
\end{align}
with homogenous Dirichlet boundary conditions for $\boldsymbol{u}$
and Neumann boundary conditions for $p$ at all boundaries. Moreover,
the boundary conditions for temperature (Figure \hyperlink{fig2}{2}) are prescribed
as follows.
\begin{align}
\frac{\partial T}{\partial x_{2}}\left(x_{1},0,\boldsymbol{\xi}\right)=\frac{\partial T}{\partial x_{2}}\left(x_{1},1,\boldsymbol{\xi}\right) & =0, & x_{1}\in\left[0,1\right],\nonumber \\
T\left(0,x_{2},\boldsymbol{\xi}\right)-T_{h}\left(x_{2},\boldsymbol{\xi}_{2}\right)=T\left(1,x_{2},\boldsymbol{\xi}\right) & =0, & x_{2}\in\left[0,1\right].
\end{align}

Moreover, $\boldsymbol{e}_{2}$ denotes $\left[\begin{array}{cc}
0 & 1\end{array}\right]^\mathbf{T}$ while $\mathrm{Pr}$ and $\mathrm{Ra}$ denote the Prandtl and Rayleigh
numbers respectively, $T_{h}$ denotes the hot-wall temperature such
that $\forall x_{2}\in$$\left[0,1\right]$,
\begin{equation}
T_{h}\left(x_{2},\boldsymbol{\xi}_{2}\right)=\bar{T}_{h}+h\left(x_{2},\boldsymbol{\xi}_{2}\right)\sin^{2}\left(\pi x_{2}\right),
\end{equation}
where $\bar{T}_{h}$ is the mean hot-wall temperature and $h$ denotes
the perturbation amplitude. 

In this study, $\mathrm{Ra}$ and $h$ are assumed to be independent
random fields, and modeled using the following Karhunen-Loeve (KL) [\hyperlink{ref35}{35}] expansions. $\forall\boldsymbol{x}\in\Omega,\boldsymbol{\xi}_{1}\in\Xi{}_{1}$,
\begin{equation}
\mathrm{Ra}\left(\boldsymbol{x},\boldsymbol{\xi}_{1}\right)=\bar{\mathrm{Ra}}+\sqrt{3}\delta_{\mathrm{Ra}}\sum_{j=1}^{s_{1}}\gamma_{\mathrm{Ra},j}\left(\boldsymbol{x}\right)\xi_{1j},
\end{equation}
where $\bar{\mathrm{Ra}}$ denotes the mean of $\mathrm{Ra}$ and
$\left\{ \xi_{1j}\sim U\left[-1,1\right]\right\}_{j=1}^{s_1} $
are $i.i.d$ random variables. Similarly, $\forall x_{2}\in\left(0,1\right),\boldsymbol{\xi}_{2}\in\Xi{}_{2}$,
\begin{equation}
h\left(x_{2},\boldsymbol{\xi}_{2}\right)=\sqrt{3}\delta_{h}\sum_{j=1}^{s_{2}}\gamma_{h,j}\left(x_2\right)\xi_{2j},
\end{equation}
where $\left\{ \xi_{2j}\sim U\left[-1,1\right]\right\}_{j=1}^{s_2} $
are $i.i.d$ random variables. Moreover, we assume that both $\mathrm{Ra}$
and $h$ have exponential kernels
\begin{align}
C_{\mathrm{Ra}}\left(\boldsymbol{x},\boldsymbol{y}\right) & =\delta_{\mathrm{Ra}}^{2}\exp\left(-\frac{\left\Vert \boldsymbol{x}-\boldsymbol{y}\right\Vert _{1}}{l_{\mathrm{Ra}}}\right),\boldsymbol{x},\boldsymbol{y}\in\Omega,\nonumber \\
C_{h}\left(x_{2},y_{2}\right) & =\delta_{h}^{2}\exp\left(-\frac{\left|x_{2}-y_{2}\right|}{l_{h}}\right),x_{2},y_{2}\in\left[0,1\right],
\end{align}
where $\delta_{\mathrm{Ra}}$, $\delta_{h}$ denote the respective
coefficient of variations and $l_{\mathrm{Ra}}$, $l_{h}$ denote
the respective correlation lengths. The analytic expressions for $\gamma_{\mathrm{Ra},j},\gamma_{h,j}:j>0$
are provided in Appendix \hyperlink{appB}{B}. Moreover, in place of the continuity equation,
the pressure Poisson equation
\begin{equation}
\boldsymbol{\nabla}^\mathbf{T}\boldsymbol{\nabla}p\left(\boldsymbol{x},\boldsymbol{\xi}\right)+\boldsymbol{\nabla}^\mathbf{T}\left(\left(\boldsymbol{u}\left(\boldsymbol{x},\boldsymbol{\xi}\right)^\mathbf{T}\boldsymbol{\nabla}\right)\boldsymbol{u}\left(\boldsymbol{x},\boldsymbol{\xi}\right)-\mathrm{Pr}\mathrm{Ra}\left(\boldsymbol{x},\boldsymbol{\xi}_{1}\right)T\left(\boldsymbol{x},\boldsymbol{\xi}\right)\boldsymbol{e}_{2}\right)=0
\end{equation}
is used to close the coupled PDE system. Table \hyperlink{tab1}{1} lists the corresponding
numerical values of the deterministic parameters used in this study.

\hypertarget{tab1}{}
\begin{table}[htbp]
\caption{Deterministic parameter values in the thermally-driven cavity
flow problem.}
\begin{center}\scriptsize
\renewcommand{\arraystretch}{1.3}
\begin{tabular}{ccccccc}
\toprule 
$\mathrm{Pr}$ & $\bar{\mathrm{Ra}}$ & $\bar{T}_{h}$ & $\delta_{\mathrm{Ra}}$ & $\delta_{h}$ & $l_{\mathrm{Ra}}$ & $l_{h}$\tabularnewline
\midrule
\midrule 
$0.71$ & $10^{3}$ & $1$ & $10$ & $0.5$ & $0.5$ & $0.5$\tabularnewline
\bottomrule
\end{tabular}
\par\end{center}
\end{table}

Each component PDE system is spatially discretized using a finite
volume method, with linear central-differencing schemes [\hyperlink{ref36}{36}],
on a uniform grid with $m\times m$ cells. Let $\boldsymbol{u}_{1}^{\prime},\boldsymbol{u}_{2}^{\prime},\boldsymbol{p}^{\prime},\boldsymbol{t}^{\prime}\in\mathbb{R}^{m^{2}}$
denote the respective vectors of cell-centroid horizontal velocity,
vertical velocity, pressure and temperature, which solve the nonlinear
system\hypertarget{eq48}{}
\begin{align}
\left(\boldsymbol{K}_{u}+\boldsymbol{A}\left(\boldsymbol{u}_{1}^{\prime},\boldsymbol{u}_{2}^{\prime}\right)\right)\boldsymbol{u}_{1}^{\prime}+\boldsymbol{B}_{1}\boldsymbol{p}^{\prime} & =\boldsymbol{0},\nonumber \\
\left(\boldsymbol{K}_{u}+\boldsymbol{A}\left(\boldsymbol{u}_{1}^{\prime},\boldsymbol{u}_{2}^{\prime}\right)\right)\boldsymbol{u}_{2}^{\prime}+\boldsymbol{B}_{2}\boldsymbol{p}^{\prime}-\boldsymbol{R}\left(\boldsymbol{\xi}_{1}\right)\boldsymbol{t}^{\prime} & =\boldsymbol{0},\nonumber \\
\boldsymbol{K}_{p}\boldsymbol{p}^{\prime}+\boldsymbol{C}_{1}\left(\boldsymbol{u}_{1}^{\prime},\boldsymbol{u}_{2}^{\prime}\right)\boldsymbol{u}_{1}^{\prime}+\boldsymbol{C}_{2}\left(\boldsymbol{u}_{1}^{\prime},\boldsymbol{u}_{2}^{\prime}\right)\boldsymbol{u}_{2}^{\prime}-\boldsymbol{S}\left(\boldsymbol{\xi}_{1}\right)\boldsymbol{t}^{\prime} & =\boldsymbol{0},\nonumber \\
\left(\boldsymbol{K}_{T}+\boldsymbol{A}\left(\boldsymbol{u}_{1}^{\prime},\boldsymbol{u}_{2}^{\prime}\right)\right)\boldsymbol{t}^{\prime}-\boldsymbol{h}\left(\boldsymbol{\xi}_{2}\right) & =\boldsymbol{0}.
\end{align}
where each term in Eq. \hyperlink{eq48}{4.8} denotes its respective discretized
operator in the coupled PDE system in Eq. \hyperlink{eq41}{4.1}. Subsequently, we formulate a modular
multi-physics setup by separating the momentum and energy components
of the coupled algebraic PDE system. As per Eq. \hyperlink{eq11}{1.1}, let $\boldsymbol{u}_{1}=\left[\boldsymbol{u}_{1}^{\prime};\boldsymbol{u}_{2}^{\prime};\boldsymbol{p}^{\prime}\right]\in\mathbb{R}^{3m^{2}}$,
$\boldsymbol{u}_{2}=\boldsymbol{t}^{\prime}\in\mathbb{R}^{m^{2}}$
denote the respective solution variables in the modular algebraic
system. The component residuals are defined as follows.
\begin{align}
\boldsymbol{f}_{1}\left(\boldsymbol{u}_{1},\boldsymbol{u}_{2},\boldsymbol{\xi}_{1}\right) & =\left[\begin{array}{cc}
\boldsymbol{K}_{u}+\boldsymbol{A}\left(\boldsymbol{u}_{1}^{\prime}\left(\boldsymbol{u}_{1}\right),\boldsymbol{u}_{2}^{\prime}\left(\boldsymbol{u}_{1}\right)\right) & \boldsymbol{0}\\
\boldsymbol{0} & \boldsymbol{K}_{u}+\boldsymbol{A}\left(\boldsymbol{u}_{1}^{\prime}\left(\boldsymbol{u}_{1}\right),\boldsymbol{u}_{2}^{\prime}\left(\boldsymbol{u}_{1}\right)\right)\\
\boldsymbol{C}_{1}\left(\boldsymbol{u}_{1}^{\prime}\left(\boldsymbol{u}_{1}\right),\boldsymbol{u}_{2}^{\prime}\left(\boldsymbol{u}_{1}\right)\right) & \boldsymbol{C}_{2}\left(\boldsymbol{u}_{1}^{\prime}\left(\boldsymbol{u}_{1}\right),\boldsymbol{u}_{2}^{\prime}\left(\boldsymbol{u}_{1}\right)\right)
\end{array}\right.\nonumber \\
 & \left.\begin{array}{c}
\boldsymbol{B}_{1}\\
\boldsymbol{B}_{2}\\
\boldsymbol{K}_{p}
\end{array}\right]\boldsymbol{u}_{1}-\left[\begin{array}{c}
\boldsymbol{0}\\
\boldsymbol{R}\left(\boldsymbol{\xi}_{1}\right)\\
\boldsymbol{S}\left(\boldsymbol{\xi}_{1}\right)
\end{array}\right]\boldsymbol{u}_{2},\nonumber \\
\boldsymbol{f}_{2}\left(\boldsymbol{u}_{1},\boldsymbol{u}_{2},\boldsymbol{\xi}_{2}\right) & =\left(\boldsymbol{K}_{T}+\boldsymbol{A}\left(\boldsymbol{u}_{1}^{\prime}\left(\boldsymbol{u}_{1}\right),\boldsymbol{u}_{2}^{\prime}\left(\boldsymbol{u}_{1}\right)\right)\right)\boldsymbol{u}_{2}-\boldsymbol{h}\left(\boldsymbol{\xi}_{2}\right).
\end{align}

The quantities of interest in this study are the statistics of the fluid velocity and temperature, and the probability density
functions (pdfs) of the kinetic energy $K$ and internal energy $E$,
defined as follows. $\forall\boldsymbol{\xi}\in\Xi$, 
\begin{equation}
K\left(\boldsymbol{\xi}\right)=\frac{1}{2}\left(\int_{\Omega}u_{1}\left(\boldsymbol{x},\boldsymbol{\xi}\right)^{2}d\boldsymbol{x}+\int_{\Omega}u_{2}\left(\boldsymbol{x},\boldsymbol{\xi}\right)^{2}d\boldsymbol{x}\right),\ E\left(\boldsymbol{\xi}\right)=\int_{\Omega}T\left(\boldsymbol{x},\boldsymbol{\xi}\right)d\boldsymbol{x}.
\end{equation}

For this numerical example, two instances (or cases) of our proposed
framework were implemented and compared against their corresponding
monolithic implementations. The block Gauss-Seidel (BGS) approach
with Newton updates in each module. In each instance, the stochastic
modules and wrappers corresponding to each implementation were developed
as $\mathtt{MATLAB}^{\mathtt{TM}}$ scripts, and tested on a 3.1 GHz
Intel i5 workstation with 4GB DDR3 memory capacity.

\subsection{Case 1: Intrusive + Intrusive}

In this instance, both modules use an intrusive propagation method
based on solving the SGS system corresponding to the respective Newton
updates to propagate the gPC coefficient matrices. In the monolithic
implementation, each module performs a projection in the global $s$-dimensional
stochastic space, while in the modular framework, each module $i$
performs a projection in its local $s_{i}$-dimensional stochastic
space. Therefore, in each module, the corresponding restriction maps
yield the modular gPC coefficient matrix samples, while the prolongation
map is defined according to the pseudospectral recovery method (Theorem
\hyperlink{thm2}{2}), using local sparse-grid quadrature rules.

For each implementation, the converged gPC coefficient matrices were
obtained for $m=20$, $s_{1}=s_{2}=4$, $p=4$ and a convergence tolerance
of $10^{-8}$ on the Newton updates. Subsequently, using these matrices
we computed the probability distribution functions of $K$ and $E$,
along with the mean and standard deviation of the fluid velocity and
temperature. The results are shown in Figures \hyperlink{fig3}{3}, \hyperlink{fig4}{4} and \hyperlink{fig5}{5} respectively.
Due to the high regularity in the solutions and consequent exponential
decay in the gPC approximation error, the results are observed to
match accurately within the prescribed tolerance of $10^{-4}$, which
was chosen according to the gPC approximation error of $1.5\times10^{-5}$
observed in the module-based hybrid framework implementation. 

\hypertarget{tab2}{}
\begin{table}[htbp]
\caption{Average mean-square errors and wall-times observed in Case
1.}
\begin{center}
\scriptsize
\renewcommand{\arraystretch}{1.3}
\begin{tabular}{cc|cc|cc|c}
\toprule 
 &  & \multicolumn{2}{c}{Monolithic} \vline & \multicolumn{2}{c}{Modular} \vline & Speedup\tabularnewline
$s_{1},s_{2}$ & $p$ & Error & Wall-time (s)  & Error & Wall-time (s) & factor\tabularnewline
\midrule
\multirow{4}{*}{$3$} & $1$ & $8.2\times10^{-2}$ & $6$ & $8.8\times10^{-2}$ & $8$ & $-0.2$\tabularnewline
\cmidrule{2-7} 
 & $2$ & $2.2\times10^{-3}$ & $27$ & $3.5\times10^{-3}$ & $13$ & $1.1$\tabularnewline
\cmidrule{2-7} 
 & $3$ & $9.3\times10^{-5}$ & $181$ & $2.1\times10^{-4}$ & $67$ & $1.7$\tabularnewline
\cmidrule{2-7} 
 & $4$ & $5.4\times10^{-6}$ & $1217$ & $1.3\times10^{-5}$ & $344$ & $2.5$\tabularnewline
\midrule 
\multirow{4}{*}{$4$} & $1$ & $9.3\times10^{-2}$ & $9$ & $1.2\times10^{-1}$ & $10$ & $-0.1$\tabularnewline
\cmidrule{2-7} 
 & $2$ & $3.3\times10^{-3}$ & $53$ & $4.7\times10^{-3}$ & $18$ & $1.9$\tabularnewline
\cmidrule{2-7} 
 & $3$ & $1.3\times10^{-4}$ & $742$ & $3.1\times10^{-4}$ & $197$ & $2.8$\tabularnewline
\cmidrule{2-7} 
 & $4$ & $6.1\times10^{-6}$ & $7345$ & $1.5\times10^{-5}$ & $1625$ & $3.5$\tabularnewline
\bottomrule
\end{tabular}
\par\end{center}
\end{table}

Subsequently, for various choices of $s_{1},s_{2},p$, we compare
the error and computation time in both implementations. In each implementation,
the error is as the average mean-square error between the gPC-based
surrogate solutions and corresponding deterministic solutions, at
$100$ random sample points. The results are shown in Table \hyperlink{tab2}{2}. We
observe that as for small error tolerances, the costs of module-based
intrusive implementation was lower than the monolithic implementation.
This can be attributed to the much faster growth in the size of SGS
systems with respect to $s$ and $p$, when the latter approach is
implemented.

The highest speedup factor for each instance of $s_1$, $s_2$ was observed at the highest order setting ($p=4$). For $s_1=s_2=3$, we observed a speedup factor of $\approx 2.5$, while for $s_1=s_2=4$, we observed a speedup factor of $approx 3.5$. We expect these gains to increase when higher values of $p$ are chosen.

\hypertarget{fig3}{}
\begin{figure}[htbp]
    \centering
    \includegraphics[bb=600bp 420bp 560bp 710bp,scale=0.59]{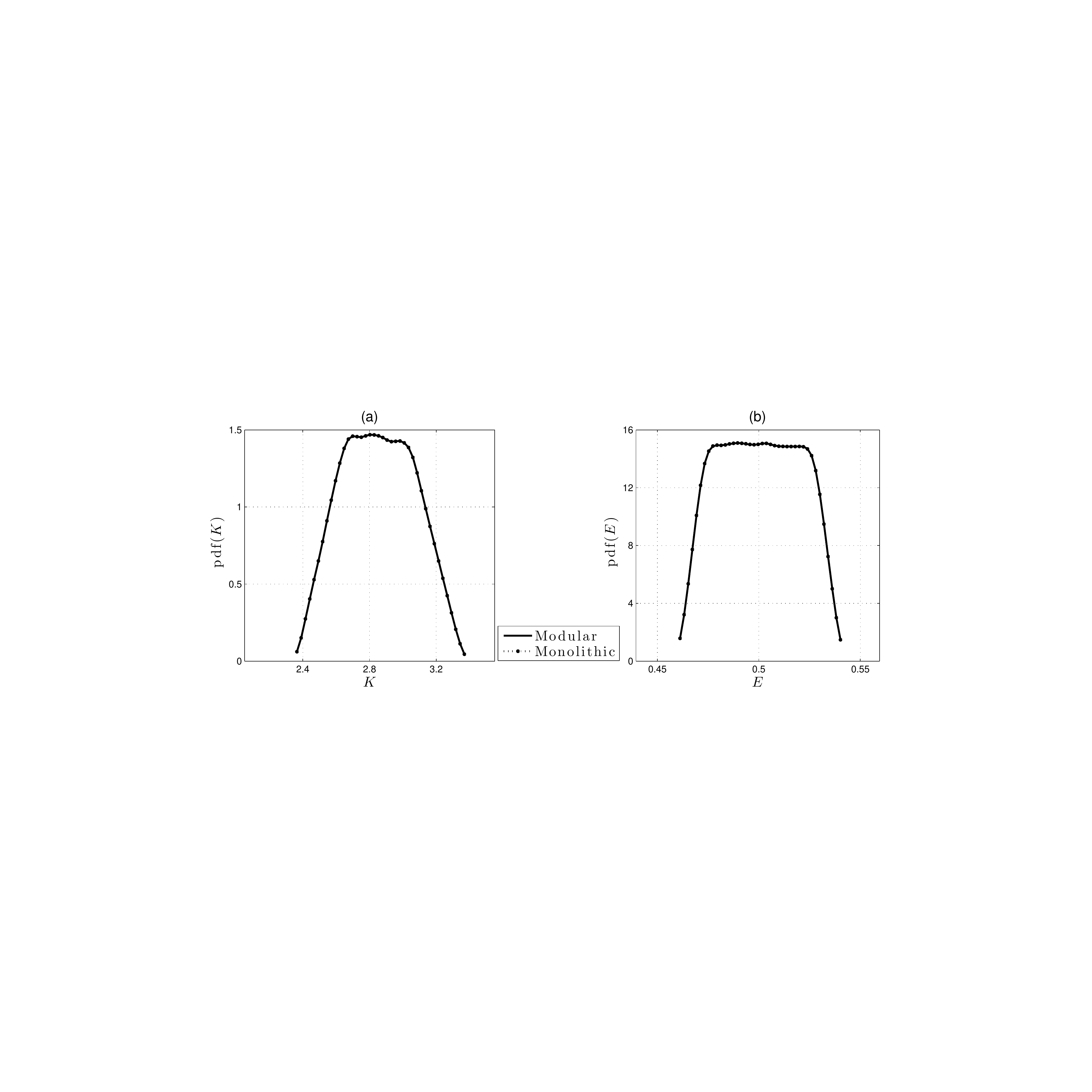}
    \caption{Probability distribution functions of fluid energies computed using both framework implementations. (a) illustrates the kinetic energy $K$ and (b) illustrates the thermal energy $E$. The KDE method was used to compute the distributions with $10^5$ samples of the gPC approximations}
    \label{fig:fig3}
\end{figure}

\hypertarget{fig4}{}
\begin{figure}[htbp]
    \centering
    \includegraphics[bb=500bp 350bp 560bp 760bp,scale=0.55]{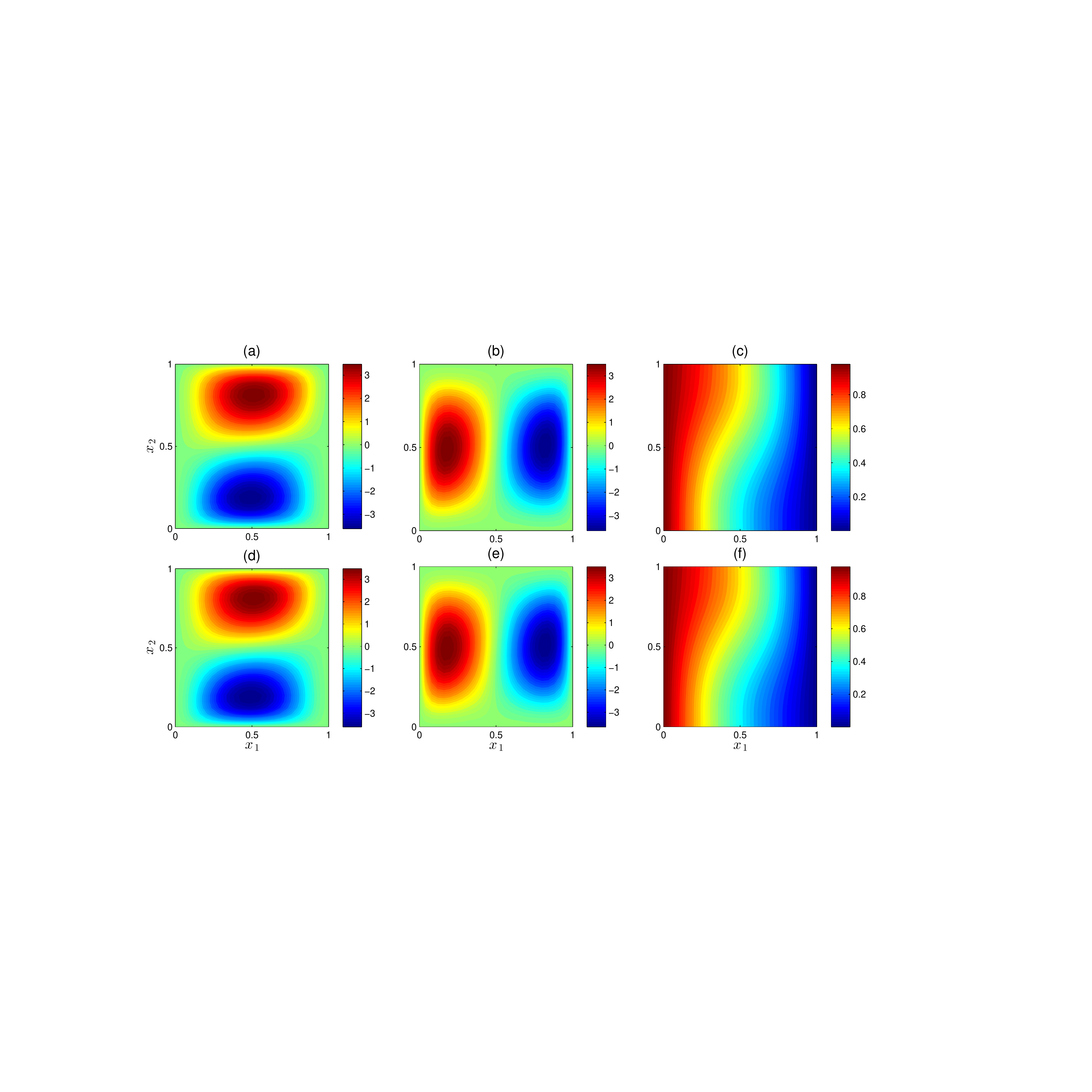}
    \caption{Mean values of solution quantities computed using both framework implementations. Subfigures (a, d), (b, e), (c, f) correspond to $u_{1}$, $u_{2}$, and  $T$ respectively, while (a, b, c) and (d, e, f) correspond to the module-based hybrid framework and monolithic framework implementations respectively.}
    \label{fig:fig4}
\end{figure}

\hypertarget{fig5}{}
\begin{figure}[htbp]
   \centering
    \includegraphics[bb=500bp 350bp 560bp 800bp,scale=0.55]{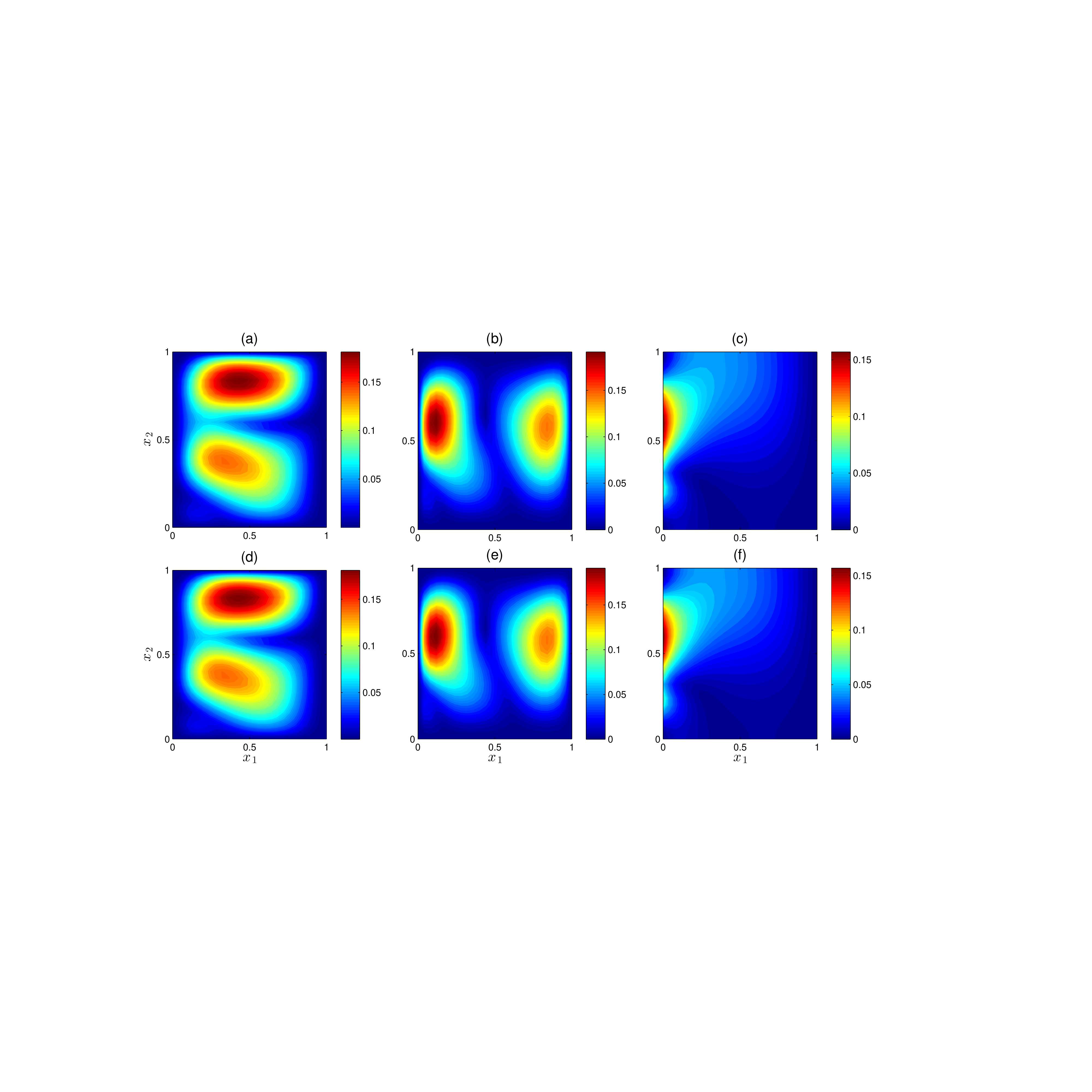}
    \caption{Standard deviation of solution quantities computed using both framework implementations. Subfigures (a, d), (b, e), (c, f) correspond to $u_{1}$, $u_{2}$, and  $T$ respectively, while (a, b, c) and (d, e, f) correspond to the module-based hybrid framework and monolithic framework implementations respectively.}

    \label{fig:fig5}
\end{figure}

\subsection{Case 2: Intrusive + Non-intrusive}

In this instance, a non-intrusive variant of module 2, based on the
pseudospectral method and corresponding global sparse-grid quadrature
rule, is implemented. Due to the nonlinearity in the underlying problem
structure, the implementation this module in the module-based hybrid
framework is the same as in the monolithic framework and therefore,
no gains in the local computational costs can be expected. However,
since module 1 is kept unchanged, we would still expect gains in overall
computational costs. 

\hypertarget{tab3}{}
\begin{table}[htbp]
\caption{Mean-square average errors and wall-times observed in Case
2.}
\begin{center}
\scriptsize
\renewcommand{\arraystretch}{1.3}
\begin{tabular}{cc|cc|cc|c}
\toprule 
 &  & \multicolumn{2}{c}{Monolithic} \vline & \multicolumn{2}{c}{Modular} \vline & Speedup\tabularnewline
$s_{1},s_{2}$ & $p$ & Error & Wall-time (s)  & Error & Wall-time (s) & factor\tabularnewline
\midrule
\multirow{4}{*}{$3$} & $1$ & $1.1\times10^{-1}$ & $6$ & $1.3\times10^{-1}$ & $7$ & $-0.1$\tabularnewline
\cmidrule{2-7} 
 & $2$ & $2.6\times10^{-3}$ & $23$ & $5.1\times10^{-3}$ & $12$ & $0.9$\tabularnewline
\cmidrule{2-7} 
 & $3$ & $1.0\times10^{-4}$ & $166$ & $2.2\times10^{-4}$ & $61$ & $1.7$\tabularnewline
\cmidrule{2-7} 
 & $4$ & $6.5\times10^{-6}$ & $1093$ & $1.6\times10^{-5}$ & $329$ & $2.3$\tabularnewline
\midrule 
\multirow{4}{*}{$4$} & $1$ & $1.3\times10^{-1}$ & $8$ & $1.5\times10^{-1}$ & $9$ & $-0.1$\tabularnewline
\cmidrule{2-7} 
 & $2$ & $4.1\times10^{-3}$ & $46$ & $6.7\times10^{-3}$ & $16$ & $1.9$\tabularnewline
\cmidrule{2-7} 
 & $3$ & $1.8\times10^{-4}$ & $671$ & $4.4\times10^{-4}$ & $185$ & $2.6$\tabularnewline
\cmidrule{2-7} 
 & $4$ & $6.8\times10^{-6}$ & $6636$ & $2.1\times10^{-5}$ & $1521$ & $3.4$\tabularnewline
\bottomrule
\end{tabular}
\par\end{center}
\end{table}

For $m=20$, $s_{1}=s_{2}=4$, $p=4$ and a convergence tolerance
of $10^{-8}$ on the Newton updates, we compared the first two solution
moments obtained using both implementations, We observed that the
results matched accurately with errors below than the prescribed tolerance
of $10^{-4}$, which was chosen according to the gPC approximation
error of $2.1\times10^{-5}$ observed in the module-based hybrid framework
implementation. When comparing these results to the results obtained
in Case 1, we once again observed an accurate match with errors below
$10^{-4}$.

As in Case 1, $s_{1},s_{2}$ and $p$ were varied for a comparison of
performance. The average mean-square error (computed using $100$
samples) and the wall-times observed in both implementations are listed
in Table \hyperlink{tab3}{3}. In comparison to Case 1, we observe a slight reduction
of costs in both implementations and slight increase in the approximation
errors. 

Moreover, since module 1 dominates the computational costs,
the gains observed in the module-based hybrid implementation were only
slightly lower than the gains observed in Case 1. Once, again, for each instance of $s_1$, $s_2$, we observed the highest computational gain at $p=4$, with higher gains expected as $p$ is further increased. For $s_1=s_2=3$, we observed a speedup factor of $\approx 2.3$, while for $s_1=s_2=4$, we observed a speedup factor of $\approx 3.4$.

\section{Conclusions and outlook}

We presented an extension of the module-based hybrid UQ framework introduced
in our previous work [1] to general nonlinear multi-physics systems.
We described the basic components of the framework, namely the restriction
and prolongation maps, which facilitates uncertainty propagation to
be abstracted down to the module level, and a seamless blending of disparate stochastic modules
for efficient global uncertainty propagation. Therefore, our proposed framework
reduces the developmental costs and overheads associated with stochastic multi-physics
modeling and simulation, when compared to a fully-coupled monolithic framework.

Besides achieving this motivating goal, we observed a speedup factor between $2.3$ and $3.5$ in
numerical experiments, where high order gPC-based propagation methods developed in our proposed framework
were compared against their respective monolithic implementations. Due to the nonlinear
structure of the model in the numerical example, the gains observed
were mainly limited to intrusive modules, where the size of the associated SGS systems were much smaller in the module-based implementation, when compared to the monolithic implementation, for the same global stochastic dimension and gPC order.

Although global multi-variate polynomials were chosen for uncertainty representation in this work, our proposed methodology can also be
demonstrated with multiresolution uncertainty propagation schemes which employ either Haar wavelets [\hyperlink{ref37}{37}] to overcome
the loss of accuracy in models exhibiting discontinuities, or multi-element gPC [\hyperlink{ref38}{38}] to overcome
the loss of accuracy with long-time integration in unsteady models.

Our experience with this framework has increased our confidence in
its viability and superior scalability of modularization over monolithic
implementation for uncertainty propagation in complex multi-physics
systems. The framework is therefore, a suitable candidate for codesign
in the next generation (exascale) of high performance computers. Efficient
strategies for parallelization in the domain of inter-module communication/data
transfer, memory manipulation are currently being investigated. Moreover,
exploring the interplay between uncertainties and numerical errors
at the modular level is also being actively investigated [\hyperlink{ref39}{39}]. 

\section*{Acknowledgement}
This research was funded by the US Department of Energy, Office of
Advanced Computing Research and Applied Mathematics Program and partially
funded by the US Department of Energy NNSA ASC Program under Contract No.
DE-AC52-07NA27344. The work was performed
as a collaboration between Stanford University and the US Department
of Energy Lawrence Livermore National Laboratory.

\newpage
\appendix
\hypertarget{appA}{}
\section*{Appendix A}

\hypertarget{lemA1}{}
\subsubsection*{Lemma A1: }

Given a matrix $\boldsymbol{A}\in\mathbb{R}^{m\times n}:m\geq n$
with $rank\left(\boldsymbol{A}\right)=n$, $\boldsymbol{B}=\left[\begin{array}{ccc}
\boldsymbol{b}_{1} & \cdots & \boldsymbol{b}_{k}\end{array}\right]\in\mathbb{R}^{m\times k}$, and an objective function $f\left(\boldsymbol{X}\right)=\left\Vert \boldsymbol{B}-\boldsymbol{A}\boldsymbol{X}\right\Vert _{F}^{2}:\mathbb{R}^{n\times k}\rightarrow\mathbb{R}$,
the minimizer of $f$ is unique and can be evaluated as follows.
\[
\boldsymbol{X}^{*}=\arg\min_{\boldsymbol{X}\in\mathbb{R}^{n\times k}}f\left(\boldsymbol{X}\right)=\left(\boldsymbol{A}^\mathbf{T}\boldsymbol{A}\right)^{-1}\boldsymbol{A}^\mathbf{T}\boldsymbol{B}.\tag{A.1}
\]

\subsubsection*{Proof: }

$\forall\boldsymbol{X}=\left[\begin{array}{ccc}
\boldsymbol{x}_{1} & \cdots & \boldsymbol{x}_{Q}\end{array}\right]\in\mathbb{R}^{n\times k}$, we have
\begin{align*}
f\left(\boldsymbol{x}\right) & =\left\Vert \boldsymbol{B}-\boldsymbol{A}\boldsymbol{X}\right\Vert _{F}^{2}\\
 & =\left\Vert \left[\begin{array}{c}
\boldsymbol{b}_{1}\\
\vdots\\
\boldsymbol{b}_{k}
\end{array}\right]-\left(\boldsymbol{I}_{k}\otimes\boldsymbol{A}\right)\left[\begin{array}{c}
\boldsymbol{x}_{1}\\
\vdots\\
\boldsymbol{x}_{k}
\end{array}\right]\right\Vert _{2}^{2}.\tag{A.2}
\end{align*}

The minimizer of $f$ can be obtained by solving ${\displaystyle \frac{\partial f}{\partial\boldsymbol{x}_{j}}=\boldsymbol{0}:1\leq j\leq k}.$
Therefore, we have 
\begin{align*}
\left[\begin{array}{c}
\boldsymbol{x}_{1}^{*}\\
\vdots\\
\boldsymbol{x}_{k}^{*}
\end{array}\right] & =\left(\left(\boldsymbol{I}_{k}\otimes\boldsymbol{A}\right)^\mathbf{T}\left(\boldsymbol{I}_{k}\otimes\boldsymbol{A}\right)\right)^{-1}\left(\boldsymbol{I}_{k}\otimes\boldsymbol{A}\right)^\mathbf{T}\left[\begin{array}{c}
\boldsymbol{b}_{1}\\
\vdots\\
\boldsymbol{b}_{k}
\end{array}\right]\\
 & =\left(\boldsymbol{I}_{k}\otimes\left(\boldsymbol{A}^\mathbf{T}\boldsymbol{A}\right)\right)^{-1}\left(\boldsymbol{I}_{k}\otimes\boldsymbol{A}^\mathbf{T}\right)\left[\begin{array}{c}
\boldsymbol{b}_{1}\\
\vdots\\
\boldsymbol{b}_{k}
\end{array}\right]\\
 & =\left(\boldsymbol{I}_{k}\otimes\left(\boldsymbol{A}^\mathbf{T}\boldsymbol{A}\right)^{-1}\right)\left(\boldsymbol{I}_{k}\otimes\boldsymbol{A}^\mathbf{T}\right)\left[\begin{array}{c}
\boldsymbol{b}_{1}\\
\vdots\\
\boldsymbol{b}_{k}
\end{array}\right]\\
 & =\left(\boldsymbol{I}_{k}\otimes\left(\left(\boldsymbol{A}^\mathbf{T}\boldsymbol{A}\right)^{-1}\boldsymbol{A}^\mathbf{T}\right)\right)\left[\begin{array}{c}
\boldsymbol{b}_{1}\\
\vdots\\
\boldsymbol{b}_{k}
\end{array}\right],\tag{A.3}
\end{align*}
which can be rewritten as follows. 
\begin{align*}
\boldsymbol{X}^{*}=\left[\begin{array}{ccc}
\boldsymbol{x}_{1}^{*} & \cdots & \boldsymbol{x}_{k}^{*}\end{array}\right] & =\left(\boldsymbol{A}^\mathbf{T}\boldsymbol{A}\right)^{-1}\boldsymbol{A}^\mathbf{T}\left[\begin{array}{ccc}
\boldsymbol{b}_{1} & \cdots & \boldsymbol{b}_{k}\end{array}\right]\\
 & =\left(\boldsymbol{A}^\mathbf{T}\boldsymbol{A}\right)^{-1}\boldsymbol{A}^\mathbf{T}\boldsymbol{B}.\tag{A.4}
\end{align*}
$\square$

\hypertarget{lemA2}{}
\subsubsection*{Lemma A2:}

If the quadrature rule $\left\{ \left(\boldsymbol{\xi}^{\left(j\right)},w^{\left(j\right)}\right) \right\} _{j=1}^{Q}$
can exactly integrate polynomials with total degree $\leq2p$ then
the gPC coefficient matrix $\hat{\boldsymbol{U}}^{p}$ obtained using
the pseudospectral formula in Eq. \hyperlink{eq211}{2.11} corresponds to the unique
stationary point of the following objective function. 
\[
f\left(\hat{\boldsymbol{Y}}\right)=\left\Vert \boldsymbol{U}-\hat{\boldsymbol{Y}}\boldsymbol{\Psi}\right\Vert _{F,\boldsymbol{W}}^{2}:\mathbb{R}^{n\times\left(P+1\right)}\rightarrow\mathbb{R},\tag{A.5}
\]
where $\forall\boldsymbol{X}=\left[\begin{array}{ccc}
\boldsymbol{x}_{1} & \cdots & \boldsymbol{x}_{Q}\end{array}\right]\in\mathbb{R}^{n\times Q}$, $\left\Vert \boldsymbol{X}\right\Vert _{F,\boldsymbol{W}}^{2}$
denotes the $\boldsymbol{W}-$weighted Frobenius pseudonorm as follows.
\[
\left\Vert \boldsymbol{X}\right\Vert _{F,\boldsymbol{W}^{Q}}=\sqrt{\left|\sum_{j=1}^{Q}w^{\left(j\right)}\left\Vert \boldsymbol{x}_{j}\right\Vert _{2}^{2}\right|}.\tag{A.6}
\]

\subsubsection*{Proof:}

$\forall\hat{\boldsymbol{Y}}=\left[\begin{array}{ccc}
\hat{\boldsymbol{y}}^{0} & \cdots & \hat{\boldsymbol{y}}^{P}\end{array}\right]\in\mathbb{R}^{n\times\text{\ensuremath{\left(P+1\right)}}}$, we have
\begin{align*}
f\left(\hat{\boldsymbol{Y}}\right) & =\left\Vert \left[\begin{array}{c}
\boldsymbol{u}^{\left(1\right)}\\
\vdots\\
\boldsymbol{u}^{\left(Q\right)}
\end{array}\right]-\left(\boldsymbol{\Psi}^\mathbf{T}\otimes\boldsymbol{I}_{n}\right)\left[\begin{array}{c}
\hat{\boldsymbol{y}}^{0}\\
\vdots\\
\hat{\boldsymbol{y}}^{P}
\end{array}\right]\right\Vert _{\boldsymbol{W}\otimes\boldsymbol{I}_{n}}^{2},\tag{A.7}
\end{align*}
where 
\[
\forall\left[\begin{array}{c}
\boldsymbol{x}_{1}\\
\vdots\\
\boldsymbol{x}_{Q}
\end{array}\right]\in\mathbb{R}^{nQ},\left\Vert \left[\begin{array}{c}
\boldsymbol{x}_{1}\\
\vdots\\
\boldsymbol{x}_{Q}
\end{array}\right]\right\Vert _{\boldsymbol{W}\otimes\boldsymbol{I}_{n}}=\left\Vert \left[\begin{array}{ccc}
\boldsymbol{x}_{1} & \cdots & \boldsymbol{x}_{Q}\end{array}\right]\right\Vert _{F,\boldsymbol{W}}.\tag{A.8}
\]
The unique stationary point of $f$ can be obtained by solving ${\displaystyle \frac{\partial f}{\partial\hat{\boldsymbol{y}}^{j}}}=\boldsymbol{0}:1\leq j\leq Q$.
Since the quadrature rule has exact accuracy for polynomials of total
degree greater than equal to $2p$, we have
\[
\sum_{j=1}^{Q}w^{\left(j\right)}\psi^{k}\left(\boldsymbol{\xi}^{\left(j\right)}\right)\psi^{l}\left(\boldsymbol{\xi}^{\left(j\right)}\right)=\delta_{kl}\Rightarrow\boldsymbol{\Psi}\boldsymbol{W}\boldsymbol{\Psi}^\mathbf{T}=\boldsymbol{I}_{P+1}.\tag{A.9}
\]
Therefore, we have
\begin{align*}
\left[\begin{array}{c}
\hat{\boldsymbol{u}}^{0}\\
\vdots\\
\hat{\boldsymbol{u}}^{P}
\end{array}\right] & =\left(\left(\boldsymbol{\Psi}^\mathbf{T}\otimes\boldsymbol{I}_{n}\right)^\mathbf{T}\left(\boldsymbol{W}\otimes\boldsymbol{I}_{n}\right)\left(\boldsymbol{\Psi}^\mathbf{T}\otimes\boldsymbol{I}_{n}\right)\right)^{-1}\\
 & \quad\times\left(\boldsymbol{\Psi}^\mathbf{T}\otimes\boldsymbol{I}_{n}\right)^\mathbf{T}\left(\boldsymbol{W}\otimes\boldsymbol{I}_{n}\right)\left[\begin{array}{c}
\boldsymbol{u}^{\left(1\right)}\\
\vdots\\
\boldsymbol{u}^{\left(Q\right)}
\end{array}\right]\\
 & =\left(\boldsymbol{\Psi}\boldsymbol{W}\boldsymbol{\Psi}^\mathbf{T}\otimes\boldsymbol{I}_{n}\right)^{-1}\left(\boldsymbol{\Psi}^\mathbf{T}\otimes\boldsymbol{I}_{n}\right)^\mathbf{T}\left(\boldsymbol{W}\otimes\boldsymbol{I}_{n}\right)\left[\begin{array}{c}
\boldsymbol{u}^{\left(1\right)}\\
\vdots\\
\boldsymbol{u}^{\left(Q\right)}
\end{array}\right]\\
 & =\left(\boldsymbol{I}_{P+1}\otimes\boldsymbol{I}_{n}\right)^{-1}\left(\left(\boldsymbol{\Psi}\boldsymbol{W}\right)\otimes\boldsymbol{I}_{n}\right)\left[\begin{array}{c}
\boldsymbol{u}^{\left(1\right)}\\
\vdots\\
\boldsymbol{u}^{\left(Q\right)}
\end{array}\right]\\
 & =\left(\left(\boldsymbol{\Psi}\boldsymbol{W}\right)\otimes\boldsymbol{I}_{n}\right)\left[\begin{array}{c}
\boldsymbol{u}^{\left(1\right)}\\
\vdots\\
\boldsymbol{u}^{\left(Q\right)}
\end{array}\right],\tag{A.10}
\end{align*}
which can be rewritten as follows.
\begin{align*}
\hat{\boldsymbol{U}}=\left[\begin{array}{ccc}
\hat{\boldsymbol{u}}^{0} & \cdots & \hat{\boldsymbol{u}}^{P}\end{array}\right] & =\left[\begin{array}{ccc}
\boldsymbol{u}^{\left(1\right)} & \cdots & \boldsymbol{u}^{\left(Q\right)}\end{array}\right]\left(\boldsymbol{\Psi}\boldsymbol{W}\right)^\mathbf{T}\\
 & =\boldsymbol{U}\boldsymbol{W}\boldsymbol{\Psi}^\mathbf{T}.\tag{A.11}
\end{align*}
As some of the entries in $\boldsymbol{W}$ may be negative, defining
$\left\Vert \cdot\right\Vert _{F,\boldsymbol{W}}^{2}$ and $\left\Vert \cdot\right\Vert _{\boldsymbol{W}\otimes\boldsymbol{I}_{n}}^{2}$
as norms would violate the strict positivity and triangular inequality
conditions. Therefore, we have instead defined them as pseudonorms.
Nonetheless, since some of the entries in $\boldsymbol{W}$ must be
positive, a unique stationary point that minimizes $f$ along some
directions would always exist. $\square$

\hypertarget{lemA3}{}
\subsubsection*{Lemma A3:}

Let $\boldsymbol{l}_{1}:\mathbb{R}^{n_{1}+n_{2}}\times\mathbb{R}^{s_{1}}\rightarrow\mathbb{R}^{n_{1}}$
and $\boldsymbol{l}_{2}:\mathbb{R}^{n_{1}+n_{2}}\times\mathbb{R}^{s_{2}}\rightarrow\mathbb{R}^{n_{2}}$
correspond to linear maps such that $\forall\left(\boldsymbol{\xi}_{1},\boldsymbol{\xi}_{2}\right)\in\Xi$,\hypertarget{eqA12}{}
\begin{align*}
\boldsymbol{u}_{1}^{\ell+1}\left(\boldsymbol{\xi}_{1},\boldsymbol{\xi}_{2}\right) & =\boldsymbol{m}_{1}\left(\boldsymbol{u}_{1}^{\ell}\left(\boldsymbol{\xi}_{1},\boldsymbol{\xi}_{2}\right),\boldsymbol{u}_{2}^{\ell}\left(\boldsymbol{\xi}_{1},\boldsymbol{\xi}_{2}\right),\boldsymbol{\xi}_{1}\right)=\boldsymbol{l}_{1}\left(\boldsymbol{y}_{1}^{\ell}\left(\boldsymbol{\xi}_{1},\boldsymbol{\xi}_{2}\right),\boldsymbol{\xi}_{1}\right),\\
\boldsymbol{u}_{2}^{\ell+1}\left(\boldsymbol{\xi}_{1},\boldsymbol{\xi}_{2}\right) & =\boldsymbol{m}_{2}\left(\boldsymbol{u}_{1}^{\ell+1}\left(\boldsymbol{\xi}_{1},\boldsymbol{\xi}_{2}\right),\boldsymbol{u}_{2}^{\ell}\left(\boldsymbol{\xi}_{1},\boldsymbol{\xi}_{2}\right),\boldsymbol{\xi}_{2}\right)=\boldsymbol{l}_{2}\left(\boldsymbol{y}_{2}^{\ell}\left(\boldsymbol{\xi}_{1},\boldsymbol{\xi}_{2}\right),\boldsymbol{\xi}_{2}\right),\tag{A.12}
\end{align*}
where $\boldsymbol{y}_{1}^{\ell}=\left[\boldsymbol{u}_{1}^{\ell};\boldsymbol{u}_{2}^{\ell}\right]$
and $\boldsymbol{y}_{2}^{\ell}=\left[\boldsymbol{u}_{1}^{\ell+1};\boldsymbol{u}_{2}^{\ell}\right]$.
The gPC coefficients of $\boldsymbol{u}_{1}^{\ell+1}$ and $\boldsymbol{u}_{2}^{\ell+1}$
can be obtained by a decomposition of the projection into ${s_{2}+p \choose p}$
and ${s_{1}+p \choose p}$ subproblems respectively.

\subsubsection*{Proof:}

We consider the first equation in Eq. \hyperlink{eqA12}{A.12} corresponding to
module 1. Each gPC coefficients of $\boldsymbol{u}_{1}^{\ell+1}$
can be obtained by projecting it in $\Xi$ against the respective
polynomial basis. Therefore, $\forall0\leq\left|\boldsymbol{j}_{1}\right|+\left|\boldsymbol{j}_{2}\right|\leq p$,
we have 
\begin{align*}
\hat{\boldsymbol{u}}_{1}^{\ell+1,\boldsymbol{j}_{1}\boldsymbol{j}_{2}} & =\int_{\mathbb{R}^{s_{2}}}\int_{\mathbb{R}^{s_{1}}}\boldsymbol{l}_{1}\left(\boldsymbol{y}_{1}^{\ell}\left(\boldsymbol{\xi}_{1},\boldsymbol{\xi}_{2}\right),\boldsymbol{\xi}_{1}\right)\psi_{1}^{\boldsymbol{j}_{1}}\left(\boldsymbol{\xi}_{1}\right)\psi_{2}^{\boldsymbol{j}_{2}}\left(\boldsymbol{\xi}_{2}\right)d\mathcal{P}_{1}\left(\boldsymbol{\xi}_{1}\right)d\mathcal{P}_{2}\left(\boldsymbol{\xi}_{2}\right)\\
 & =\int_{\mathbb{R}^{s_{2}}}\int_{\mathbb{R}^{s_{1}}}\left(\boldsymbol{l}_{1}\left(\sum_{\left|\boldsymbol{k}_{2}\right|=0}^{p}\sum_{\left|\boldsymbol{k}_{1}\right|=0}^{p-\left|\boldsymbol{k}_{2}\right|}\boldsymbol{y}_{1}^{\ell,\boldsymbol{k}_{1}\boldsymbol{k}_{2}}\psi_{1}^{\boldsymbol{k}_{1}}\left(\boldsymbol{\xi}_{1}\right)\psi_{2}^{\boldsymbol{k}_{2}}\left(\boldsymbol{\xi}_{2}\right),\boldsymbol{\xi}_{1}\right)\psi_{1}^{\boldsymbol{j}_{1}}\left(\boldsymbol{\xi}_{1}\right)\right.\\
 & \left.\times\psi_{2}^{\boldsymbol{j}_{2}}\left(\boldsymbol{\xi}_{2}\right)d\mathcal{P}_{1}\left(\boldsymbol{\xi}_{1}\right)d\mathcal{P}_{2}\left(\boldsymbol{\xi}_{2}\right)\right)\\
 & =\sum_{\left|\boldsymbol{k}_{2}\right|=0}^{p}\sum_{\left|\boldsymbol{k}_{1}\right|=0}^{p-\left|\boldsymbol{k}_{2}\right|}\left(\int_{\mathbb{R}^{s_{2}}}\int_{\mathbb{R}^{s_{1}}}\boldsymbol{l}_{1}\left(\hat{\boldsymbol{y}}_{1}^{\ell,\boldsymbol{k}_{1}\boldsymbol{k}_{2}},\boldsymbol{\xi}_{1}\right)\psi_{1}^{\boldsymbol{j}_{1}}\left(\boldsymbol{\xi}_{1}\right)\psi_{1}^{\boldsymbol{k}_{1}}\left(\boldsymbol{\xi}_{1}\right)\psi_{2}^{\boldsymbol{j}_{2}}\left(\boldsymbol{\xi}_{2}\right)\right.\\
 & \left.\times\psi_{2}^{\boldsymbol{k}_{2}}\left(\boldsymbol{\xi}_{2}\right)d\mathcal{P}_{1}\left(\boldsymbol{\xi}_{1}\right)d\mathcal{P}_{2}\left(\boldsymbol{\xi}_{2}\right)\right)\\
 & =\int_{\mathbb{R}^{s_{1}}}\boldsymbol{l}_{1}\left(\sum_{\left|\boldsymbol{k}_{1}\right|=0}^{p-\left|\boldsymbol{j}_{2}\right|}\hat{\boldsymbol{y}}_{1}^{\ell,\boldsymbol{k}_{1}\boldsymbol{j}_{2}}\psi_{1}^{\boldsymbol{k}_{1}}\left(\boldsymbol{\xi}_{1}\right),\boldsymbol{\xi}_{1}\right)\psi_{1}^{\boldsymbol{j}_{1}}\left(\boldsymbol{\xi}_{1}\right)d\mathcal{P}_{1}\left(\boldsymbol{\xi}_{1}\right).\tag{A.13}
\end{align*}
Therefore, to evaluate the gPC coefficients $\left\{ \hat{\boldsymbol{u}}_{1}^{\ell+1,\boldsymbol{j}_{1}\boldsymbol{j}_{2}}:0\leq\left|\boldsymbol{j}_{1}\right|+\left|\boldsymbol{j}_{2}\right|\leq p\right\} $,
we would only require the gPC coefficients of $\boldsymbol{u}_{1}^{\ell}$
and $\boldsymbol{u}_{2}^{\ell}$ with indices belonging to the set that can be written as
$\left\{ \boldsymbol{k_{1}}\boldsymbol{j}_{2}\in\mathbb{N}_{0}^{s_{1}}\times\mathbb{N}_{0}^{s_{2}}:0\leq\left|\boldsymbol{k}_{1}\right|+\left|\boldsymbol{j}_{2}\right|\leq p\right\} $.
Since these sets are disjoint, we can decompose the projection integrals
into independent subproblems corresponding to different values of
$\boldsymbol{j}_{2}$. 

The same procedure can be followed to prove the result for the second
equation in Eq. \hyperlink{eqA12}{A.12}. $\square$

\hypertarget{lemA4}{} 
\subsubsection*{Lemma A4:}

Let $\boldsymbol{A}_{1}:\mathbb{R}^{s_{1}}\rightarrow\mathbb{R}^{n_{1}\times n_{1}}$
and $\boldsymbol{b}_{1}:\mathbb{R}^{n_{1}+n_{2}}\times\mathbb{R}^{s_{1}}\rightarrow\mathbb{R}^{n_{1}}$
be the random invertible matrix and linear forcing vector derived
from module opreator $\boldsymbol{m}_{1}$. Also, let $\boldsymbol{A}_{2}:\mathbb{R}^{s_{2}}\rightarrow\mathbb{R}^{n_{2}\times n_{2}}$
and $\boldsymbol{b}_{2}:\mathbb{R}^{n_{1}+n_{2}}\times\mathbb{R}^{s_{2}}\rightarrow\mathbb{R}^{n_{2}}$
be the random invertible matrix and linear forcing vector derived
from module opreator $\boldsymbol{m}_{2}$. Therefore, $\forall\left(\boldsymbol{\xi}_{1},\boldsymbol{\xi}_{2}\right)\in\Xi$,\hypertarget{eqA14}{}
\begin{align*}
\boldsymbol{A}_{1}\left(\boldsymbol{\xi}_{1}\right)\boldsymbol{u}_{1}^{\ell+1}\left(\boldsymbol{\xi}_{1},\boldsymbol{\xi}_{2}\right) & =\boldsymbol{A}_{1}\left(\boldsymbol{\xi}_{1}\right)\boldsymbol{m}_{1}\left(\boldsymbol{u}_{1}^{\ell}\left(\boldsymbol{\xi}_{1},\boldsymbol{\xi}_{2}\right),\boldsymbol{u}_{2}^{\ell}\left(\boldsymbol{\xi}_{1},\boldsymbol{\xi}_{2}\right),\boldsymbol{\xi}_{1}\right)\\
 & =\boldsymbol{b}_{1}\left(\boldsymbol{y}_{1}^{\ell}\left(\boldsymbol{\xi}_{1},\boldsymbol{\xi}_{2}\right),\boldsymbol{\xi}_{1}\right),\\
\boldsymbol{A}_{2}\left(\boldsymbol{\xi}_{2}\right)\boldsymbol{u}_{2}^{\ell+1}\left(\boldsymbol{\xi}_{1},\boldsymbol{\xi}_{2}\right) & =\boldsymbol{A}_{2}\left(\boldsymbol{\xi}_{2}\right)\boldsymbol{m}_{2}\left(\boldsymbol{u}_{1}^{\ell+1}\left(\boldsymbol{\xi}_{1},\boldsymbol{\xi}_{2}\right),\boldsymbol{u}_{2}^{\ell}\left(\boldsymbol{\xi}_{1},\boldsymbol{\xi}_{2}\right),\boldsymbol{\xi}_{2}\right)\\
 & =\boldsymbol{b}_{2}\left(\boldsymbol{y}_{2}^{\ell}\left(\boldsymbol{\xi}_{1},\boldsymbol{\xi}_{2}\right),\boldsymbol{\xi}_{2}\right),\tag{A.14}
\end{align*}
where $\boldsymbol{y}_{1}^{\ell}=\left[\boldsymbol{u}_{1}^{\ell};\boldsymbol{u}_{2}^{\ell}\right]$
and $\boldsymbol{y}_{2}^{\ell}=\left[\boldsymbol{u}_{1}^{\ell+1};\boldsymbol{u}_{2}^{\ell}\right]$.
The stochastic Galerkin system (SGS) associated with the gPC coefficients
of $\boldsymbol{u}_{1}^{\ell+1}$ and $\boldsymbol{u}_{2}^{\ell+1}$
can be decomposed into ${s_{2}+p \choose p}$ and ${s_{1}+p \choose p}$
subproblems respectively.

\subsubsection*{Proof:}

We consider the the first equation in Eq. \hyperlink{eqA14}{A.14} and its corresponding
SGS. The left hand matrix of this system would have ${s+p \choose p}\times{s+p \choose p}$
submatrix blocks of size $n_{1}\times n_{1}$ each. The $\left(\boldsymbol{j}_{1}\boldsymbol{j}_{2},\boldsymbol{k}_{1}\boldsymbol{k}_{2}\right)$-th
block can be evaluated as follows.\hypertarget{eqA15}{}
\begin{align*}
 & \int_{\mathbb{R}^{s_{2}}}\int_{\mathbb{R}^{s_{1}}}\boldsymbol{A}_{1}\left(\boldsymbol{\xi}_{1}\right)\psi_{1}^{\boldsymbol{j}_{1}}\left(\boldsymbol{\xi}_{1}\right)\psi_{2}^{\boldsymbol{j}_{2}}\left(\boldsymbol{\xi}_{2}\right)\psi_{1}^{\boldsymbol{k}_{1}}\left(\boldsymbol{\xi}_{1}\right)\psi_{2}^{\boldsymbol{k}_{2}}\left(\boldsymbol{\xi}_{2}\right)d\mathcal{P}_{1}\left(\boldsymbol{\xi}_{1}\right)d\mathcal{P}_{2}\left(\boldsymbol{\xi}_{2}\right)\\
 & =\begin{cases}
\int_{\mathbb{R}^{s_{1}}}\boldsymbol{A}_{1}\left(\boldsymbol{\xi}_{1}\right)\psi_{1}^{\boldsymbol{j}_{1}}\left(\boldsymbol{\xi}_{1}\right)\psi_{1}^{\boldsymbol{k}_{1}}\left(\boldsymbol{\xi}_{1}\right)d\mathcal{P}_{1}\left(\boldsymbol{\xi}_{1}\right) & \boldsymbol{j}_{2}=\boldsymbol{k}_{2}\\
\boldsymbol{0} & \boldsymbol{j}_{2}\neq\boldsymbol{k}_{2}
\end{cases}.\tag{A..15}
\end{align*}
The right hand vector would have ${s+p \choose p}$ subvector blocks
of size $n_{1}$ each. The $\left(\boldsymbol{j}_{1}\boldsymbol{j}_{2}\right)$-th
block can be evaluated using Theorem 1 as follows.\hypertarget{eqA16}{}
\begin{align*}
 & \int_{\mathbb{R}^{s_{2}}}\int_{\mathbb{R}^{s_{1}}}\boldsymbol{l}_{1}\left(\boldsymbol{y}_{1}^{\ell}\left(\boldsymbol{\xi}_{1},\boldsymbol{\xi}_{2}\right),\boldsymbol{\xi}_{1}\right)\psi_{1}^{\boldsymbol{j}_{1}}\left(\boldsymbol{\xi}_{1}\right)\psi_{2}^{\boldsymbol{j}_{2}}\left(\boldsymbol{\xi}_{2}\right)d\mathcal{P}_{1}\left(\boldsymbol{\xi}_{1}\right)d\mathcal{P}_{2}\left(\boldsymbol{\xi}_{2}\right)\\
 & =\int_{\mathbb{R}^{s_{1}}}\boldsymbol{l}_{1}\left(\sum_{\left|\boldsymbol{k}_{1}\right|=0}^{p-\left|\boldsymbol{j}_{2}\right|}\hat{\boldsymbol{y}}_{1}^{\ell,\boldsymbol{k}_{1}\boldsymbol{j}_{2}}\psi_{1}^{\boldsymbol{k}_{1}}\left(\boldsymbol{\xi}_{1}\right),\boldsymbol{\xi}_{1}\right)\psi_{1}^{\boldsymbol{j}_{1}}\left(\boldsymbol{\xi}_{1}\right)d\mathcal{P}_{1}\left(\boldsymbol{\xi}_{1}\right).\tag{A.16}
\end{align*}
Therefore, the block diagonal structure of the left hand matrix, as
indicated by Eq. \hyperlink{eqA15}{A.15}, and the independence of each subvector
in the right hand vector, as indicated by Eq. \hyperlink{eqA16}{A.16}, is used
for decomposing the SGS into smaller subsystems of linear equations.
In each subsystem, for various values of $\boldsymbol{j}_{2}\in\mathbb{N}_{0}^{s_{2}}$,
we can independently compute $\left\{ \hat{\boldsymbol{u}}_{1}^{\ell+1,\boldsymbol{k}_{1}\boldsymbol{j}_{2}}:0\leq\left|\boldsymbol{k}_{1}\right|\leq p-\left|\boldsymbol{j}_{2}\right|\right\} $.
The same procedure can be followed to prove the result for the second
equation  in Eq. \hyperlink{eqA14}{A.14}. $\square$

\newpage
\hypertarget{appB}{}
\section*{Appendix B}

\subsection*{Karhunen-Loeve expansion for the exponential kernel}

Given an $n$-dimensional spatial domain $\Omega\subseteq\mathbb{R}^{n}$,
let $C_{u}:\Omega\rightarrow\mathbb{R}^{+}$ denote the exponential
covariance kernel of a spatially varying random quantity $u:\Omega\rightarrow\mathbb{R}$.
Therefore, $C_{u}:\forall\boldsymbol{x}=\left[\begin{array}{ccc}
x_{1} & \cdots & x_{n}\end{array}\right]^\mathbf{T},\boldsymbol{y}=\left[\begin{array}{ccc}
y_{1} & \cdots & y_{n}\end{array}\right]^\mathbf{T}\in\Omega$, 
\[
C_{u}\left(\boldsymbol{x},\boldsymbol{y}\right)=\exp\left(-\frac{\left\Vert \boldsymbol{x}-\boldsymbol{y}\right\Vert _{1}}{l}\right)=\prod_{j=1}^{n}\exp\left(-\frac{\left|x_{j}-y_{j}\right|}{l}\right),\tag{B.1}
\]
where $l$ denotes the correlation length. Subsequently, we can define
the KL expansion of $u$ using an infinite set of random variables
$\left\{ \xi_{\boldsymbol{j}}:\boldsymbol{j}\in\mathbb{N}^{n}\right\} $
as follows. $\forall\boldsymbol{x}\in\Omega$, \hypertarget{eqB2}{}

\begin{align*}
u\left(\boldsymbol{x}\right)-\bar{u}\left(\boldsymbol{x}\right) & =\sum_{\boldsymbol{j}\in\mathbb{R}^{n}}\gamma_{\boldsymbol{j}}\left(\boldsymbol{x}\right)\xi_{\boldsymbol{j}}\\
 & =\sum_{j_{1}\in\mathbb{R}}\cdots\sum_{j_{n}\in\mathbb{R}}\gamma_{j_{1}\ldots j_{n}}\left(\boldsymbol{x}\right)\xi_{j_{1}\ldots j_{n}}\\
 & =\sum_{j_{1}\in\mathbb{R}}\cdots\sum_{j_{n}\in\mathbb{R}}\prod_{k=1}^{n}g_{j_{k}}\left(r_{k}\right)\xi_{j_{1}\ldots j_{n}}\tag{B.2}
\end{align*}
where $\forall j>0,$ if $\zeta_{j}$ solves
\[
l\zeta_{j}+\tan\left(\frac{\zeta_{j}}{2}\right)=0,\tag{B.3}
\]
and $\zeta_{j+1}>\zeta_{j}>0$, then $\forall x\in\mathbb{R},$ 
\[
g_{j}\left(x\right)=\begin{cases}
2{\displaystyle \sqrt{\frac{l\zeta_{j}}{1+l^{2}\zeta_{j}^{2}}}\frac{\cos\left(\zeta_{j}x\right)}{\sqrt{\zeta_{j}+\sin\left(\zeta_{j}\right)}}} & j\ \mathrm{is\ odd},\\
2{\displaystyle \sqrt{\frac{l\zeta_{j}}{1+l^{2}\zeta_{j}^{2}}}\frac{\sin\left(\zeta_{j}x\right)}{\sqrt{\zeta_{j}-\sin\left(\zeta_{j}\right)}}} & j\ \mathrm{is\ even}.
\end{cases}\tag{B.4}
\]
Therefore, as is required in $\S$\hyperlink{sec4}{4}, a truncated KL expansion can
be easily obtained from the single index form of the expansion in
Eq. \hyperlink{eqB2}{B.2}.
\end{document}